%% file: GlobalFit.tex
\documentclass[a4paper,11pt]{article}
\pdfoutput=1
\usepackage{jheppub}
\usepackage{slashed,subcaption,amssymb,amsmath,bm}
\usepackage[utf8]{inputenc}
\usepackage{amsfonts}
\usepackage{bbold}
\usepackage{mathtools}
\usepackage{pdflscape}
\usepackage[dvipsnames]{xcolor}
\usepackage{forloop}
\usepackage{arydshln}

\usepackage{ifpdf}
\graphicspath{{./img/}}
\usepackage{hyperref}
\usepackage{cleveref}
\usepackage{tikz}
\usepackage{float}
\usepackage{cancel}
\usepackage{booktabs}
\usepackage{multirow}
\usepackage{rotating}
\usepackage{soul}

\newcommand{\gev}{\text{ GeV}}
\newcommand{\tev}{\text{ TeV}}

\newcommand{\Tr}[1]{\text{Tr}\left[#1\right]}
\newcommand{\brackets}[1]{\left(#1\right)}
\newcommand{\sqbrackets}[1]{\left[#1\right]}

\newcommand{\appropto}{\mathrel{\vcenter{
  \offinterlineskip\halign{\hfil$##$\cr
    \propto\cr\noalign{\kern2pt}\sim\cr\noalign{\kern-2pt}}}}}

\def\figureautorefname~#1\null{Fig.\,#1\null}
\def\tableautorefname~#1\null{Tab.\,#1\null}

\def\equationautorefname~#1\null{Eq.\,(#1)\null}

\DeclareUnicodeCharacter{2212}{-}

\makeindex

\makeatletter\g@addto@macro\bfseries\boldmath
\makeatother

\title{Extending Global Fits of 4D Composite Higgs Models with Partially Composite Leptons}

\author[a]{Ethan Carragher,}
\author[b]{Kenn Goh,}
\author[c]{Wei Su,}
\author[b]{Martin White,}
\author[b]{and Anthony G. Williams}

\affiliation[a]{Rudolf Peierls Centre for Theoretical Physics, University of Oxford, Parks Road, Oxford OX1 3PU, United Kingdom}
\affiliation[b]{ARC Centre of Excellence for Dark Matter Particle Physics, Department of Physics, University of Adelaide, South Australia 5005, Australia}
\affiliation[c]{School of Science, Shenzhen Campus of Sun Yat-sen University, No. 66, Gongchang Road, \\ Guangming District, Shenzhen, Guangdong 518107, P.R. China }

\emailAdd{ethan.carragher@physics.ox.ac.uk}
\emailAdd{kennshern.goh@adelaide.edu.au}
\emailAdd{suwei26@mail.sysu.edu.cn}
\emailAdd{martin.white@adelaide.edu.au}
\emailAdd{anthony.williams@adelaide.edu.au,}

\abstract{
We perform the first convergent Bayesian global fits of 4D Composite Higgs Models with partially-composite third generation quarks and leptons based on the minimal $SO(5) \rightarrow SO(4)$ symmetry breaking pattern. We consider two models with the $\tau$ lepton and its associated neutrino in different representations of $SO(5)$. Fitting each model with a wide array of experimental constraints allows us to analyse the Bayesian evidence and currently-observed fine-tuning of each model by calculating the Kullback-Leibler divergence between their respective priors and posteriors. Notably both models are found to be capable of satisfying all constraints simultaneously at the $3\sigma$ level at scales of $< 5$ TeV. From a Bayesian viewpoint of naturalness the model with leptons in the $\mathbf{14}$ and $\mathbf{10}$ representations is preferred over those in the $\mathbf{5}$ representation due to its lower fine-tuning. Finally, we consider the experimental signatures for the preferred parameters in these models, including lepton partner decay signatures and gluon-fusion produced Higgs signal strengths, and discuss their potential phenomenology at future high-luminosity LHC runs.
}

\keywords{Technicolor and Composite Models, Beyond Standard Model, Effective Field Theories, Global Symmetries, Naturalness, Composite Higgs}

\date{}

\begin{document}

\maketitle

\input{./tex/Introduction}

\input{./tex/Model}

\input{./tex/Scanning}

\input{./tex/Results}
\input{./tex/Pheno}

\section{Conclusions}
\label{sec:conclusions}

In this work we have extended previous global fits of two-site Minimal 4D Composite Higgs Models by including composite partners for the third-generation leptons in various representations of $SO(5)$, and imposing relevant constraints on the added lepton parameters. We considered all quark partners as embedded in the $\mathbf{5}$ representation due to its favourable phenomenology, and chose to embed the respective partners of the left-handed lepton doublet and right-handed tau lepton into the $\mathbf{5}-\mathbf{5}$ and $\mathbf{14}-\mathbf{10}$ representations (giving the so-called LM4DCHM$^{5-5-5}_{5-5}$ and LM4DCHM$^{5-5-5}_{14-10}$ models) to study how these might affect the fine-tuning and viability of the theory.

Our fits revealed that, while both models are capable of satisfying all imposed constraints, and hence are equally viable from a frequentist perspective, the LM4DCHM$^{5-5-5}_{14-10}$ is clearly preferred from a Bayesian viewpoint over the LM4DCHM$^{5-5-5}_{5-5}$, having the greater Bayesian evidence by several orders of magnitude. This conclusion is quite robust against both the choice of prior distribution (uniform or logarithmic) and the prior bounds imposed on the parameter spaces. For all choices of prior we used, the LM4DCHM$^{5-5-5}_{14-10}$ had the favourable posterior-averaged log-likelihood, owing primarily to the SM mass and oblique constraints which it can more easily satisfy. However, the main factor favouring the LM4DCHM$^{5-5-5}_{14-10}$ is its lower fine-tuning as measured by the Kullback-Leibler divergence between the prior and posterior distributions. The exact fine-tuning of the models is strongly prior dependent, although always in favour of the LM4DCHM$^{5-5-5}_{14-10}$ for the priors we considered. This is to be expected from theoretical considerations, since the LM4DCHM$^{5-5-5}_{5-5}$ suffers from a double-tuning necessary to achieve a viable Higgs potential. Hence, the inclusion of lepton partners is seen to have a significant effect on the fine-tuning of the model, despite the relatively low compositeness of the $\tau$ compared to third generation quarks.

We also analysed the phenomenology of the two models in their viable regions of parameter space. The phenomenology of the composite gauge bosons and quark partners remains unchanged from what was found in the quark-only M4DCHM$^{5-5-5}$ in Ref.~\cite{Ethan}, but now there are also heavy lepton partners. These lepton partners are generically predicted to lie well above $1$~TeV, with the most probable decay channels being $N_4 \rightarrow W \tau$, $L_4 \rightarrow H \tau$, $L_4 \rightarrow Z \tau$ and $E2 \rightarrow W \tau$. In both models, the lightest composite lepton tended to be $L_4$. In the LM4DCHM$^{5-5-5}_{14-10}$, this predominantly decays through both the $H \tau$ and $Z \tau$ channels, whereas in the LM4DCHM$^{5-5-5}_{5-5}$, it decays almost always into a $Z \tau$ pair. However, conservative estimates of the production cross sections of the heavy leptons revealed that only in a small fraction of the parameter space could any heavy lepton resonance of the LM4DCHM$^{5-5-5}_{14-10}$ possibly be discovered at the HL-LHC, while those of the LM4DCHM$^{5-5-5}_{5-5}$ all lie far beyond the reach of the HL-LHC.

Predictions for the Higgs signal strengths for both models generally align well with experimental bounds. However, more natural values of the NGB decay constant $f \lesssim 1.25$~TeV are currently slightly disfavoured by current experimental measurements compared to higher values. When compared to the Higgs phenomenology of the quark-only M4DCHM$^{5-5-5}$, values $2.5$~TeV~$\lesssim f \lesssim 3.5$~TeV are disfavoured as a result of including the third-generation leptons.

While the results presented in this paper clearly indicate that the LM4DCHM$^{5-5-5}_{5-5}$ model is less natural and more fine-tuned, it is important to note that the model does have parameter sets that fit the data. Naturalness plays a central role in CHM searches, but it is only by experiment that models can be ruled out. Current prospects for the precision of Higgs signal strength measurements at the HL-LHC \cite{ATLAS:2018jlh,CMS-PAS-FTR-16-002} promise to provide effective tests of these models.

It would be interesting to consider possible modifications to the framework considered here, such as alternative quark embeddings in LM4DCHMs, as both Ref.~\cite{BarnardFT} and this work have shown promising results for \textbf{14} representation leptons. Furthermore, there may also be room for including right-handed neutrinos, which are well-motivated due to their ability to provide a potential dark matter candidate and to explain the observed baryonic asymmetry \cite{Ahmadvand_2021}. The complexity of the resulting parameter spaces, however, may make it difficult to obtain consistent convergent results.

\section{Acknowledgements}

We thank Peter Stangl for allowing us to use \texttt{pypngb} for this work. MJW, AGW and KG are funded by the ARC Centre of Excellence for Dark Matter Particle Physics CE200100008 and are further supported by the Centre for the Subatomic Structure of Matter (CSSM). EC is supported by the Clarendon Fund Scholarship in partnership with the Oxford-Berman Graduate Scholarship. WS is supported by the Natural Science Foundation of China (NSFC) under grant number 12305115 and the Shenzhen Science and Technology Program (Grant No. 202206193000001, 20220816094256002) 

\appendix

\input{./tex/Appendix}

\newpage
\bibliographystyle{JHEP}
\bibliography{tex/refs}

\end{document}

%% file: tex/Introduction.tex
\section{Introduction}
\label{introduction}

The hierarchy problem continues to motivate the development of beyond-Standard Model physics models. If the Standard Model (SM) couples to any new physics at high energies, then the SM Higgs boson should receive contributions to its mass of the same order as those energy scales, which raises the question of why the physical Higgs mass is comparatively small. Unless the new physics is structured in such a way that the various mass contributions cancel each other out, this observation can only be explained within the framework of the SM by an enormous fine-tuning of the bare Higgs mass parameter. We concern ourselves here with a popular alternative framework, the Composite Higgs Model (CHM), in which the Higgs boson is a bound state of some new strongly-interacting ``composite" sector at the few-TeV scale. This protects the Higgs from large mass corrections and so removes the need for much fine-tuning~\cite{kaplan1984,kaplan1984b,kaplan1985}.

The minimal viable CHM incorporates the four Higgs doublet fields as the pseudo-Nambu-Goldstone bosons (pNGBs) of spontaneous $SO(5) \rightarrow SO(4)$ symmetry breaking within the composite sector~\cite{contino2003,agashe2005}. The other SM fields arise as ``partially composite" superpositions of elementary and composite fields. Masses are communicated to the SM particles through their composite components, so heavier SM particles are generically expected to interact more strongly with the composite sector~\cite{kaplan1991}. It is common to realise this scenario as a multi-site model (the low-energy effective 4D theory of a 5D Randall-Sundrum model~\cite{ArkaniHamed:2001ca,Hill:2000mu}), since the Higgs effective potential is finite and calculable in this case. These so-called Minimal 4D CHMs (M4DCHMs) have been the subject of considerable research~\cite{contino2003,agashe2005,contino2006,Contino:2006nn,contino2007b,giudice2007,contino2007,Panico:2007qd}.

One of the primary objectives of research into M4DCHMs is to understand which features lead to more natural models. Early studies in this direction were largely concerned with the fine-tuning effects of the partially composite top and bottom quarks~\cite{de2012,panico2012}, neglecting lighter SM particles due to their weaker interactions with the composite sector. There it was understood that embedding the composite quark partners in the $\mathbf{5}$ or $\mathbf{10}$ representations of $SO(5)$ would require precise cancellations to happen in the Higgs potential to produce electroweak symmetry breaking, whereas this ``double tuning" would not be present if using the $\mathbf{14}$ representation~\cite{panico2012,matsedonskyi2012}.

M4DCHMs have since been investigated more comprehensively through detailed numerical scans~\cite{Carena,Niehoff:2015iaa,BarnardCC,BarnardFT,Ethan}. In particular, our previous work Ref.~\cite{Ethan} included the first convergent global fits of such models with partially composite third-generation quarks. There it was found that embedding all of the composite quark partners in the $\mathbf{5}$ representation (the so-called M4DCHM$^{5-5-5}$ model) is preferred from both an experimental and Bayesian standpoint despite its double tuning, since this was the only model that predicted realistic $H \rightarrow \gamma \gamma$ signal strengths. A similar conclusion was found earlier in Ref.~\cite{Carena}. This raises the interesting question of whether including the composite third-generation lepton partners in the M4DCHM$^{5-5-5}$ could lead to a more attractive model by further reducing its fine-tuning, depending on their representations~\cite{carmona2015}.

In the present work we tackle this question by extending the fits of Ref.~\cite{Ethan} to include the partial compositeness of the $\tau$ lepton and its associated neutrino\footnote{Truly consistent realisations of the M4DCHM would include partial compositeness of all SM particles. However, thorough exploration of the resulting high-dimensional parameter space is currently intractable.}. We will be fitting two different models: one in which the partners of the third-generation lepton doublet and those of the right-handed $\tau$ are respectively embedded in the $\mathbf{5}-\mathbf{5}$ combination of representations, and one in which they are embedded in the $\mathbf{14}-\mathbf{10}$ representations, within the (leptonic) M4DCHM$^{5-5-5}$. Following the notation of Ref.~\cite{DanielsThesis}, we denote these models as the LM4DCHM$^{5-5-5}_{5-5}$ and the LM4DCHM$^{5-5-5}_{14-10}$. We adopt the successful fitting method of Ref.~\cite{Ethan} and use the nested sampling algorithm \texttt{PolyChord}~\cite{Handley_2015,Handley:2015fda} to fit our models to all of the previous experimental constraints, including SM masses, electroweak precision tests, Z decay ratios, Higgs signal strengths, and composite quark partner mass bounds, as well as the newly applicable constraints of the $\tau$ mass and lepton partner mass bounds. We will further use our fit results to examine the branching ratios of the heavy lepton partners, and shed more light on the prior-dependence of our previous results. 

The question of how partially-composite leptons impact fine-tuning has in fact already been investigated to an extent in Ref.~\cite{BarnardFT}, using a novel measure of fine-tuning that captures double- and higher-order tuning effects. It was found that including lepton partners in the $\mathbf{14}$ representation indeed reduces the fine-tuning significantly, as might be expected. Our analysis differs from Ref.~\cite{BarnardFT} in that we use a larger variety of constraints, impose rigorous statistical convergence, and we interpret the fine-tuning in a Bayesian sense as the Kullback-Leibler divergence from the prior to the posterior distribution on parameter space. This notion of fine-tuning, along with the Bayesian evidence, allows the models to be judged by their naturalness in the spirit of the Higgs mass problem that CHMs aim to solve.

This paper is organised as follows. The models we will be fitting are specified in \Cref{model}, and we detail our scanning method in \Cref{scanning_method}. We present our fit results and compare the two models in \Cref{results}, before analysing their expected experimental signatures in \Cref{exp_sig_section}. We present conclusions in \Cref{sec:conclusions}.

%% file: tex/Model.tex
\section{Model overview}
\label{model}

The models we will be using in this work fall into the class of two-site leptonic M4DCHMs, which have been discussed extensively in the literature. As such, we will provide only a brief overview of the models here and refer the reader to previous treatments for further details~\cite{BarnardFT,DanielsThesis,Panico:2015jxa}. The first site contains elementary fields with the same quantum numbers as the fields of the Standard Model (excluding the Higgs), which we will denote with superscript zeros, while the second site contains the new composite fields and the Higgs. Each site is acted upon by a separate $G \equiv SU(3)_C \times SO(5) \times U(1)_X$ symmetry so that the elementary and composite fields mix into SM fields with the correct colour, weak isospin, and hypercharge representations. The subgroup of $SO(5)$ locally isomorphic to $SU(2)_{L} \times SU(2)_{R}$ allows the electric charge $E =T^3_L + Y$ to be defined in the usual way, where here the hypercharge is $Y = T^3_R + X$ and $T^{3}_{L,R}$ are the conventional third generators of $SU(2)_{L,R}$. The overall product group $G^{1}{\times}G^{2}$ spontaneously breaks to its diagonal subgroup $G^{1+2}$, giving rise to NGBs $\Omega_i$ that link the two sites together, while a further $SO(5) \rightarrow SO(4)$ symmetry breaking produces pNGBs that are identified as the fields of the Higgs doublet.

\subsection{Boson sector}
\label{boson sector}
The bosonic sector of the two-site LM4DCHM is entirely fixed by the symmetry structure. The Lagrangian contains the elementary and composite gauge fields, with contributions from the NGBs:
\begin{align}
    \mathcal{L}_{\text{boson}} = & -\frac{1}{4}\text{Tr}[G^0_{\mu\nu}G^{0\mu\nu}] -\frac{1}{4}\text{Tr}[W^0_{\mu\nu}W^{0\mu\nu}] -\frac{1}{4}B^0_{\mu\nu}B^{0\mu\nu} && \left.\vphantom{\frac{1}{4}}\right\rbrace \text{ elementary} \nonumber \\
    & -\frac{1}{4}\text{Tr}[\rho_{G\mu\nu}\rho^{\mu\nu}_G] -\frac{1}{4}\text{Tr}[\rho_{\mu\nu}\rho^{\mu\nu}] -\frac{1}{4}\rho_{X\mu\nu}\rho_X^{\mu\nu} && \left.\vphantom{\frac{1}{4}}\right\rbrace \text{ composite}\nonumber \\
    & +\sum_{i = 1,X,G}\frac{f^2_i}{4}Tr[(D_\mu\Omega_i)^\dagger(D^\mu\Omega_i)] + \frac{f^2_2}{2}(D_\mu\Omega_2\Phi_0)^\dagger(D^\mu\Omega_2\Phi_0), &\text{[NGB]}
\end{align}
where $G^0_{\mu\nu},W^0_{\mu\nu}$ and $B^0_{\mu\nu}$ are field strength tensors of the form
\begin{equation}
    A_{\mu\nu} =\partial_\mu A_\nu - \partial_\nu A_\mu +ig[A_\mu, A_\nu]
\end{equation}
that define the kinetic terms for the elementary $SU(3)_c \times SU(2)_{L} \times U(1)_Y$ gauge fields, and likewise, $\rho_{G\mu\nu},\rho_{\mu\nu}$ and $\rho_{X\mu\nu}$ the kinetic terms for the composite $SU(3)_c \times SO(5) \times U(1)_X$ gauge fields. The composite sector introduces ten massive vector bosons for $SO(5)$, eight ``heavy gluons" for $SU(3)$, and one massive abelian $U(1)_X$ resonance. We take the vacuum vector that breaks $SO(5)^{1} \rightarrow SO(4)$ to be $\Phi_0 = (0,0,0,0,1)^{\intercal}$.

As for the NGB contributions, $\Omega_i$ ($i = 1,2,X,G$) are matrices that parameterise the symmetry breakings and transform under appropriate bifundamental representations:
\begin{align}
\begin{array}{cccc}
    SO(5)^0 \times SO(5)^1 :& \Omega_1 \rightarrow g_0 \Omega_1 g^{-1}_1, \quad \ & U(1)^0_X \times U(1)^1_X :& \Omega_X \rightarrow g_0 \Omega_X g^{-1}_1,\\[0.1cm]
    SO(5)^1 \times SO(4)\hphantom{{}^1} :& \Omega_2 \rightarrow g_1 \Omega_2 h^{-1}, \quad \ & SU(3)^0_C \times SU(3)^1_C :& \Omega_G \rightarrow g_0 \Omega_G g^{-1}_1,
\end{array}
\end{align}
where $g_a$ denotes transformations from Site $a$, and $h \in SO(4)$. The decay constants $f_i$ in the NGB terms correspond to the scales of these symmetry breakings. Most NGBs are unphysical and can be gauged away, with the sole exception of the Higgs field, which is parameterised in the product $\Omega := \Omega_1 \Omega_2$. Because of this, it has an associated symmetry breaking scale $f$ given by
\begin{align}
    \frac{1}{f^{2}} = \frac{1}{f^{2}_{1}} + \frac{1}{f^{2}_{2}}.
\label{decay_constants_relation}
\end{align}
This is related to the Higgs vev $v$ by
\begin{align}
    f \equiv \frac{v}{s_{\langle h \rangle}} = \frac{246}{s_{\langle h \rangle}} \gev,
\label{Eq:f_vev_relation}
\end{align}
where $s_{\langle h \rangle}$ is the misalignment of the vacuum states. This parameter plays a crucial role in our scans as it roughly defines the energy scale of new composite physics as well as the na{\"i}ve energy cutoff of the model $\Lambda_f=4\pi f $.

\subsection{Fermion sector}

From this point on, we will be labelling our models with the notation LM4DCHM$^{q-t-b}_{\ell-\tau}$. Superscripts $q-t-b$ specify the $SO(5)$ representations under which the composite partners of the elementary $q^0_L = (t^0_L, b^0_L)^\intercal$, $t^0_R$, $b^0_R$ fields respectively transform; similarly, $\ell-\tau$ subscripts specify the $SO(5)$ representations of the composite partners for the elementary leptons $\ell^0_L = (\nu^0_L, \tau^0_L)^\intercal$, $\tau^0_R$. Note that we do not include right-handed neutrinos in our models, even though doing so can lead to a nice realisation of the type-III seesaw mechanism~\cite{carmona2015} and help with reproducing the observed baryon asymmetry \cite{Ahmadvand_2021}, because the resulting increase in dimensionality of our parameter spaces would require more computational resources than were feasible in the present study.

We fix the quark partners to lie in the fundamental $\mathbf{5}$ representation, as results from previous studies suggest the M4DCHM$^{5-5-5}$ to be the least fine-tuned amongst other quark-only models \cite{Carena,Ethan}. We will then be considering two possible combinations of the lepton partner representations, constituting the LM4DCHM$^{5-5-5}_{5-5}$ and LM4DCHM$^{5-5-5}_{14-10}$ models. The corresponding Lagrangians are simply
\begin{align}
    \mathcal{L}_\text{fermion} &= \mathcal{L}_\text{elem. quark} + \mathcal{L}_\text{elem. lepton} + \mathcal{L}_\text{comp. quark} + \mathcal{L}_\text{comp. lepton} \nonumber \\
                          &= \mathcal{L}_\text{elem. quark} + \mathcal{L}_\text{elem. lepton} +
                          \mathrlap{\mathcal{L}^{5-5-5}}
                           \hphantom{\mathcal{L}_\text{comp. quark}}
 + \mathcal{L}_{l-\tau}.
\end{align}
The covariant derivatives of the elementary fields are the same as those of their SM counterparts, with elementary coupling strengths $g_{0}$, $g'_{0}$, and $g^{0}_{s}$, while the composite partners have covariant derivatives
\begin{align}
D_\mu \Psi &= \left(\partial_\mu  - i g_\rho \rho^A_\mu T^A - i g_X \rho_X X  - i g_G \rho^a_{G_{\mu}} \frac{\lambda^a}{2} \right) \Psi.
\end{align}
Here there are new composite gauge couplings $g_{\rho}$, $g_{X}$, and $g_{G}$ corresponding to the $SO(5)$, $U(1)_{X}$, and $SU(3)_{C}$ gauge fields. The $\lambda^{a}$ are generators of $SU(3)_{C}$ that ensure the quark partners all couple to the heavy gluons with strength $g_{G}$ and the lepton partners do not couple at all.

The composite quark sector is the same across our models, containing composite partner multiplets $\Psi^{t}$ and $\tilde{\Psi}^{t}$ in the $\mathbf{5}$ representation of $SO(5)$ and having $+2/3$ $U(1)_{X}$ charge, and $\Psi^{b}$ and $\tilde{\Psi}^{b}$ in the $\mathbf{5}$ representation of $SO(5)$ and having $-1/3$ $U(1)_{X}$ charge. In total the Lagrangian is
\begin{align}
\mathcal{L}^{5-5-5}_\text{comp. quark} =\  &\bar{q}^0_L i \slashed{D} q^0_L + \bar{t}^0_R i \slashed{D} t^0_R + \bar{b}^0_R i \slashed{D} b^0_R && \left.\vphantom{\bar{b}^0_R \slashed{D}}\right\rbrace \text{ elementary} \nonumber \\[0.1cm]
& + \bar{\Psi}^t\left(i \slashed{D} - m_{t} \right)\Psi^t + \bar{\tilde{\Psi}}^t \left( i\slashed{D} - m_{\tilde{t}} \right) \tilde{\Psi}^t && \left.\vphantom{\bar{\tilde{\Psi}}^u}\right\rbrace \text{ composite} \nonumber\\[0.1cm]
& +  \Delta_{tL} \bar{\psi}^t_L \Omega_1 \Psi^t_R +  \Delta_{tR} \bar{\psi}^t_R \Omega_1 \tilde{\Psi}^t_L && \left.\vphantom{\bar{\psi}^u_R}\right\rbrace \text{ link} \nonumber\\[0.1cm]
& - m_{Y_t} \bar{\Psi}^t_L \tilde{\Psi}^t_R - Y_t \bar{\Psi}^t_L \Phi \Phi^\dagger \tilde{\Psi}^t_R && \left.\vphantom{\bar{\psi}^u_R}\right\rbrace \text{ Yukawa}\nonumber \\[0.1cm]
& + (t \rightarrow b) + \text{h.c.} \label{eq:fundamental_fermion_lagrangian_5-5-5}
\end{align}
The first two lines here are standard kinetic terms for the fields, with the composite fields having Dirac masses $m_{t}$ and $m_{\tilde{t}}$ (and $m_{b}$ and $m_{\tilde{b}}$). The third line shows mixing between the elementary fields and their composite partners with strengths $\Delta_{tL,tR}$ (and $\Delta_{bL,bR}$). The fourth line shows an off-diagonal mass mixing term between the composite partners with strength $m_{Y_{t}}$ and a Yukawa-like term between them with strength $Y_{t}$. The multiplets $\psi$ contain the elementary quarks furnishing incomplete representations of $SO(5)$,
\begin{align}
\psi^t_L = \frac{1}{\sqrt{2}} \left( \begin{matrix}
    b^0_L \\[1pt]
    -ib^0_L \\[1pt]
    t^0_L \\[1pt]
    it^0_L \\[1pt]
    0
\end{matrix} \right), \quad \psi^t_R  = \left( \begin{matrix}
        {} \\[1pt]
         \vec{0}  \\[1pt]
        {} \\[1pt]
        {} \\[1pt]
        \hline
        t^0_R
    \end{matrix} \right), \quad \psi^b_L = \frac{1}{\sqrt{2}} \left( \begin{matrix}
    t^0_L\\[1pt]
    i t^0_L\\[1pt]
    - b^0_L\\[1pt]
    i b^0_L\\[1pt]
    0
\end{matrix} \right), \quad \psi^b_R  = \left( \begin{matrix}
        {} \\[1pt]
         \vec{0}  \\[1pt]
        {} \\[1pt]
        {} \\[1pt]
        \hline
        b^0_R
    \end{matrix} \right),
\end{align}
which therefore \textit{explicitly} break the $SO(5)$ symmetry when mixing with the composite multiplets.

\subsection{LM4DCHM$^{5−5−5}_{5-5}$}
The composite lepton sector of the LM4DCHM$^{5−5−5}_{5-5}$ is very similar to the composite quark Lagrangian in \Cref{eq:fundamental_fermion_lagrangian_5-5-5}, except that here we need only two composite multiplets $\Psi^{\tau}$ and $\tilde{\Psi}^{\tau}$ in the $\mathbf{5}$ of $SO(5)$ with $U(1)_{X}$ charges $-1$. The corresponding Lagrangian is
\begin{align}
\mathcal{L}_{5-5} = \ &\bar{l}_L^0i\slashed{D} l_L^0 + \bar{\tau}^0_Ri\slashed{D}\tau^0_R && \left.\vphantom{\bar{\tau}^0_R}\right\rbrace \text{ elementary}\nonumber\\
                                 &+ \bar{\Psi}^{\tau}(i\slashed{D}-m_{\tau})\Psi^{\tau}+\bar{\Tilde{\Psi}}^{\tau}(i\slashed{D}-m_{\Tilde{{\tau}}})\Tilde{\Psi}^{\tau} && \left.\vphantom{\bar{\Tilde{\Psi}}^{\tau}}\right\rbrace \text{ composite}\nonumber\\
                                 &+ \Delta_{\tau L}\bar{\psi^\tau_L}\Omega_1\Psi^{\tau}_R + \Delta_{\tau R}\bar{\psi^\tau_R}\Omega_1\Tilde{\Psi}^{\tau}_L && \left.\vphantom{\bar{\Tilde{\Psi}}^{\tau}}\right\rbrace \text{ link}\nonumber\\
                                 &- m_{Y\tau}\bar{\Psi}^{\tau}_L\Tilde{\Psi}^{\tau}_R - Y_\tau\bar{\Psi}^{\tau}_L\Phi\Phi^\dagger\Tilde{\Psi}^{\tau}_R, && \left.\vphantom{\bar{\Tilde{\Psi}}^{\tau}}\right\rbrace \text{ Yukawa}
\label{eq:Llepton_55}
\end{align}
where the elementary leptons fit into incomplete representations of $SO(5)$ as
\begin{equation}
\label{eq:tauf5}
    \psi^\tau_L = \frac{1}{\sqrt{2}}\begin{pmatrix}\nu_L \\ i\nu_L \\ -\tau_L \\ i\tau_L \\ 0 \end{pmatrix}, \ \ \ \psi^\tau_R = \begin{pmatrix}  0 \\ 0 \\ 0 \\ 0 \\ \tau_R \end{pmatrix}.
\end{equation}
Here only two partner multiplets are needed because only the $\tau$ needs a mass.

\subsection{LM4DCHM$^{5−5−5}_{14-10}$}
Here the composite partner $\Psi^{\tau}$ transforms in the traceless symmetric $\mathbf{14}$ representation of $SO(5)$, while $\tilde{\Psi}^{\tau}$ transforms in the antisymmetric $\mathbf{10}$. Both transform adjointly ($\Psi \rightarrow g \Psi g^{-1}$), and have zero $U(1)_{X}$ charge. The general gauge-invariant Lagrangian is then

\begin{align}
\label{eq:Llepton_1410}
\mathcal{L}_{14-10} = \ &\bar{l}_L^0i\slashed{D} l_L^0 + \bar{\tau}^0_Ri\slashed{D}\tau^0_R \nonumber+\Tr{\bar{\Psi}^{\tau}(i\slashed{D}-m_{\tau})\Psi^{\tau}}+\Tr{\bar{\Tilde{\Psi}}^{\tau}(i\slashed{D}-m_{\Tilde{{\tau}}})\Tilde{\Psi}^{\tau}} \nonumber\\
&+ \Delta_{\tau L}\Tr{\bar{\psi^\tau_L}\Omega_1\Psi^{\tau}_R\Omega_1^\dagger} + \Delta_{\tau R}\Tr{\bar{\psi^\tau_R}\Omega_1\Tilde{\Psi}^{\tau}_L\Omega_1^\dagger} \nonumber - Y_\tau\Phi^\dagger\bar{\Psi}^l_L\tilde{\Psi}^\tau_R\Phi
\end{align}

where the elementary leptons have been put into the incomplete representations

\begin{align}
    \psi^\tau_L &= \frac{1}{2} \left( \begin{array}{c|c}
        {} & i \tau^0_L\\[1.5pt]
        0_{4{\times}4} &   \tau^0_L\\[1.5pt]
        {} & i \nu^0_L\\[1.5pt]
        {} & - \nu^0_L\\[1.5pt]
        \hline
        i \tau^0_L \ \ \tau^0_L \ \ i \nu^0_L \ - \nu^0_L & 0\\
    \end{array} \right), \quad \psi^\tau_R  = \frac{\tau^0_R}{\sqrt{8}}  \left(     \begin{matrix}
        0  & 0  & i & -1 & 0\\[1pt]
        0  & 0  & 1 & i  & 0\\[1pt]
        -i & -1 & 0 & 0  & 0\\[1pt]
        1  & -i & 0 & 0  & 0\\[1pt]
        0  & 0  & 0 & 0  & 0    
    \end{matrix} \right),
\end{align}

See \cite{Ethan} for explicit forms of the $\mathbf{14}$ and $\mathbf{10}$ partner multiplets. Note that in this representation, there will be heavy leptons of electric charge $-1$, including $SU(2)_{L}$ singlets of mass $m_{\tau}$, $m_{\tilde{\tau}}$, and $\sqrt{m^{2}_{\tilde{\tau}} + \Delta^{2}_{\tau R}}$, $SU(2)_L$ doublets of mass $M_{\pm}(m_{\tau},m_{\tilde{\tau}},Y_{\tau}/2,\Delta_{\tau L})$ and $M_{\pm}(m_{\tau},m_{\tilde{\tau}},Y_{\tau}/2,0)$, and $SU(2)_L$ triplets with masses $m_{\tau}$ and $m_{\tilde{\tau}}$, where
\begin{equation}
    M^2_\pm(x_1, x_2, x_3, x_4):= \frac{|\vec{x}|^2}{2}\pm \sqrt{\frac{|\vec{x}|^2}{4} - (x_1^2x_2^2 + x_2^2x_4^2 + x_3^2x_4^2)}.
\end{equation}
This can be seen from the singular values of the mass matrices presented in~\Cref{appendix_mass_matrices}.

\subsection{Higgs potential}
\label{Higgs_potential}
We will introduce the calculation of the quantum effective potential for the Higgs boson starting from the low-energy effective fermionic Lagrangian, since this is computationally efficient and sufficiently accurate at the energy ranges we are probing. In this approximation the heavy composite fermions are integrated out, leading to their elementary fermion counterparts having an effective Lagrangian
\begin{align}
    \mathcal{L}^{\text{eff}}_{\text{comp. fermions}} = \sum_{\psi = t,b,\nu,\tau}&[\bar{\psi}^0_L \slashed{p}(1 + \Pi_{\psi L}(p^2))\psi^0_L + \bar{\psi}^0_R \slashed{p}(1 + \Pi_{\psi R}(p^2))\psi^0_R \\ \nonumber
    &+ \bar{\psi}^0_LM_\psi(p^2)\psi^0_R + h.c.]
\end{align}
for model-dependent form factors $\Pi_\psi$ and $M_\psi$, whose exact forms are given in \Cref{appendix_correlators}. 

In terms of these functions, the fermionic contribution to the Higgs potential is
\begin{align}
    V_{\text{fermion}}^{\text{eff}}(h)= -2 \sum_{\psi=t,b,\tau,\nu} N_\psi \int \frac{\text{d}p^2_E}{16 \pi^2} \ln\left[(1+\Pi_{\psi L}(-p^2_E))(1+\Pi_{\psi R}(-p^2_E)) - \frac{|M_\psi (-p^2_E)|^2}{p^2_E}\right]
\label{eq:Higgs_potentialFF}
\end{align}
where $p_{E}$ is the Euclidean momentum, and $N_t=N_b=3$ and $N_\tau=N_\nu=1$ are the colour factors of the fields. 

For the Higgs potential expanded in $s_h := \sin(h/f)$,
\begin{equation}
    V(h):=\gamma^2 s^2_h + \beta^4 s^4_h,
\end{equation}
we can find the coefficients $\gamma$ and $\beta$ and then calculate the Higgs vev $\langle h \rangle$ through
\begin{equation}
    s_{\langle h \rangle} = \frac{\gamma}{2 \beta}
\end{equation}
and the Higgs mass is then 

\begin{equation}
   m_H:=\sqrt{8 \beta(1-s^2_{\langle h \rangle})}\frac{s_{\langle h \rangle}}{f}.
\end{equation}

Gauge boson contributions to $\gamma$ are calculated analytically \cite{BarnardFT},
\begin{align}
    \gamma_{\text{gauge}} = - \frac{9 m^{4}_{\rho} \brackets{m^{2}_{a} - m^{2}_{\rho}} t_{\theta}}{64 \pi^{2} \brackets{m^{2}_{a} - \brackets{1 + t_{\theta}} m^{2}_{\rho}}} \ln\sqbrackets{\frac{m^{2}_{a}}{\brackets{1 + t_{\theta}} m^{2}_{\rho}}}
\end{align}
to first order in $t_{\theta} := g_{0}/g_{\rho}$. Here we have made use of the masses of the lightest composite gauge bosons $m_{\rho}$ and $m_{a}$, defined by
\begin{align}
    m_{\rho}^{2} := \frac{1}{2} g_{\rho}^{2} f_{1}^{2}, \qquad m_{a}^{2} := \frac{1}{2} g_{\rho}^{2} (f_{1}^{2} + f_{2}^{2}).
\label{eq:gauge_boson_mass_parameters_definitions}
\end{align}
Gauge boson contributions to $\beta$ are neglected.

%% file: tex/Scanning.tex
\section{Scanning method}
\label{scanning_method}
In this section we detail the exact specifications of our scans, including the scanning algorithm, the parameterisation and priors we use to scan over the model parameters, and the constraints applied to the models. Results of the scans are presented in \Cref{results}.

\subsection{Scan algorithm}
Our scans make use of the nested sampling algorithm \texttt{PolyChord} \cite{Handley_2015}. Nested sampling takes an initial distribution of parameter points and iteratively discards the lowest likelihood points (defined below), replacing them with new points of higher likelihood. Sampled points which have not been discarded are termed ``live points". There is a fixed number $n_{\text{live}}$ of live points in each iteration. \texttt{PolyChord} uses the strategy of slice sampling, which we have found allows us to comprehensively explore the multi-dimensional parameter space, as well as determine potential multi-modal posteriors in a computationally efficient fashion. Furthermore, nested sampling provides an estimate of the Bayesian evidence for each model, which facilitates model comparison.

Points $\mathbf{p}$ in the parameter space of a given model $\mathcal{M}$ are sampled according to a prior distribution $\pi (\textbf{p}|\mathcal{M})$, which is imposed by hand. The priors we use are specified below. To each point we assign a likelihood value
\begin{align}
    \mathcal{L}(\mathbf{p}) = e^{-\frac{1}{2}\chi^2(\mathbf{p})},
\label{eq:likelihood_chi2_definition}
\end{align}
where $\chi^2(\mathbf{p})$ is the total chi-squared associated to the point, given our constraints. Constraints $i$ which are uncorrelated with any other contribute an amount
\begin{align}
    \chi^2_i(\mathbf{p}) = \frac{( O^{\text{theo}}_i(\mathbf{p}) - O^{\text{exp}}_i )^2}{\sigma^2_i},
\end{align}
where $O^{\text{theo}}(\mathbf{p})$ and $O^{\text{exp}}$ are respectively the predicted and experimental values of the given observable, and $\sigma_i$ its total (theoretical and experimental) uncertainty. For observables that are correlated, we instead use
\begin{align}
    \chi^2(\mathbf{p}) = ( \mathbf{\mathcal{O}}^{\text{theo}}(\mathbf{p}) - \mathbf{\mathcal{O}}^{\text{exp}} )^\intercal C^{-1} ( \mathbf{\mathcal{O}}^{\text{theo}}(\mathbf{p}) - \mathbf{\mathcal{O}}^{\text{exp}} )
\end{align}
where the correlated observables have been vectorised and $C$ is the covariance matrix that encompasses their theoretical and experimental uncertainties.

The posterior distribution over the parameter space is then given by Bayes' Theorem as
\begin{equation}
    P(\textbf{p}|\mathcal{M}) = \frac{\mathcal{L}(\textbf{p})\pi(\textbf{p}|\mathcal{M})}{Z(\mathcal{M})}\, ,
\end{equation}
where
\begin{equation}
\label{eq:Z}
    Z(\mathcal{M}) = \int d\mathbf{p}\ \mathcal{L}(\textbf{p})\pi(\textbf{p}|\mathcal{M})
\end{equation}
is the Bayesian evidence for the model $\mathcal{M}$. The evidence is a single number that quantifies the favourability of the model $\mathcal{M}$ from a Bayesian perspective, balancing how well it can fit experiment (which is the sole measure of favourability from a frequentist viewpoint) with its naturalness. This interpretation is made precise by the relation
\begin{align}
    \ln (\mathcal{Z}) = \langle \ln(\mathcal{L}) \rangle_{P} - D_{KL},
\label{Eq:lnZ}
\end{align}
where $\langle \ln(\mathcal{L}) \rangle_{P}$ is the log-likelihood averaged over the posterior distribution, and
\begin{align}
    D_{KL} = \int d\mathbf{p} \ P(\mathbf{p}|\mathcal{M}) \ln \left(\frac{P(\mathbf{p}|\mathcal{M})}{\pi(\mathbf{p}|\mathcal{M})} \right)
\end{align}
is the Kullback-Leibler (KL) divergence, which measures the difference between the prior and posterior distributions, i.e. the fine-tuning of the model. However, note that this measure of fine-tuning is prior dependent.

\texttt{PolyChord} estimates the evidence by the Riemann sum
\begin{align}
    Z(\mathcal{M}) \approx \sum_{i} (X_{i-1} - X_{i}) \mathcal{L}_{i},
\end{align}
where $X_{i}$ is the prior-weighted volume of the live points in iteration $i$, and $\mathcal{L}_{i}$ is the smallest likelihood among all of the live points in that iteration. Since points are sampled from the prior distribution, the prior-weighted volumes at each iteration are approximated as $X_{i} \approx [n_{\text{live}}/(n_{\text{live}}+1)]^{i}$. The error associated to these approximations is discussed in Ref.~\cite{Handley_2015}. We deem our scans to have converged when the posterior mass remaining in the most recent iteration $i$ of live points, $Z_{\text{live}} \approx X_{i} \langle \mathcal{L} \rangle_{\text{live}}$, falls below $10^{-3}$ times the value of the evidence calculated from all of the previous iterations.

\subsection{Scan parameters}
\label{scan_parameters}

\begin{table}[h]
\begin{center}
\begin{tabular}{l|l|l}
LM4DCHM & \multicolumn{1}{c}{\(\textbf{5}-\textbf{5}\)} & \multicolumn{1}{|c}{\(\textbf{14}-\textbf{10}\)} \\
\midrule
Decay constants &  $f$, $f_1$, $f_X$, $f_G$ & $f$, $f_1$, $f_X$, $f_G$ \\[2pt]
Gauge couplings &  $g_\rho$, $g_X$, $g_G$ & $g_\rho$, $g_X$, $g_G$ \\[2pt]
\midrule
Quark link couplings & $\Delta_{t_L}$, $\Delta_{t_R}$, $\Delta_{b_L}$, $\Delta_{b_R}$ & $\Delta_{t_L}$, $\Delta_{t_R}$, $\Delta_{b_L}$, $\Delta_{b_R}$  \\[2pt]
Quark on-diagonal masses & $m_{t}$, $m_{\tilde{t}}$, $m_{b}$, $m_{\tilde{b}}$ & $m_{t}$, $m_{\tilde{t}}$, $m_{b}$, $m_{\tilde{b}}$ \\[2pt]
Quark off-diagonal masses & $m_{Y_t}$, $m_{Y_b}$ & $m_{Y_t}$, $m_{Y_b}$ {} \\[2pt]
Quark proto-Yukawa couplings & $Y_t$, $Y_b$ & $Y_t$, $Y_b$ \\
\midrule
Lepton link couplings & $\Delta_{\tau_L}$, $\Delta_{\tau_R}$ &  $\Delta_{\tau_L}$, $\Delta_{\tau_R}$  \\[2pt]
Lepton on-diagonal masses & $m_{\tau}$, $m_{\tilde{\tau}}$ & $m_{\tau}$, $m_{\tilde{\tau}}$  \\[2pt]
Lepton off-diagonal masses & $m_{Y_\tau}$ &   {} \\[2pt]
Lepton proto-Yukawa couplings & $Y_\tau$  & $Y_\tau$ \\
\midrule
Dimensionality & \multicolumn{1}{c}{25} & \multicolumn{1}{|c}{24} 
\end{tabular}
\end{center}
\caption{Parameters present in each model.}
\label{tab:model_parameters}
\end{table}

The full list of Lagrangian parameters in each of our models is provided in~\Cref{tab:model_parameters} for convenience. These are not the parameters we scan over exactly; instead, we choose to scan over the gauge boson masses $m_{\rho}$ and $m_{a}$ from~\Cref{eq:gauge_boson_mass_parameters_definitions} directly in place of the decay constants $f$ and $f_{1}$. We also employ the approach of Refs.~\cite{Ethan,BarnardCC,BarnardFT,Carena} and scan over all mass-dimension parameters in units of $f$, only fixing $f$ afterwards by \Cref{Eq:f_vev_relation}. This step significantly reduces the computational expense of performing comprehensive scans.

\begin{table}
\begin{center}
\begin{tabular}{ @{}lllc @{} }
\toprule
Model & Parameters & Scan Range & Prior\\
\midrule
\multirow{15}{*}{Both} & $m_{\rho}/f,\ m_{a}/f$ & $[1/\sqrt{2}, 4 \pi]$ & \multirow{4}{*}{Uniform}\\[8pt]
 & $f_{X}/f,\ f_{G}/f$ & $[0.5, 2 \sqrt{3}]$\\[4pt]
 & $g_{\rho},\ g_{X},\ g_{G}$ & $[1.0, 4\pi]$\\[3pt]
 \cdashline{2-4} \\[-10pt]
 {}& $\Delta_{t_{L}}/f$ & $[e^{-0.25}, e^{1.5}]$ & \multirow{12}{*}{Logarithmic}\\[4pt]
 & $\Delta_{t_{R}}/f$ & $[e^{-0.75}, 4 \pi]$\\[4pt]
 & $\Delta_{b_{L}}/f$ & $[e^{-5.0}, e^{-3.0}]$\\[4pt]
 & $\Delta_{b_{R}}/f$ & $[e^{-0.5}, 4 \pi]$\\[4pt]
 & $m_{t}/f,\ m_{\tilde{b}}/f$ & $[e^{-0.5}, e^{1.5}]$\\[4pt]
 & $m_{\tilde{t}}/f$ & $[e^{-1.0}, 4 \pi]$\\[4pt]
 & $m_{b}/f$ & $[e^{-1.0}, e^{1.5}]$\\[4pt]
 & $m_{Y_{t}}/f$ & $[e^{-8.5}, 4 \pi]$\\[4pt]
 & $m_{Y_{b}}/f$ & $[e^{-0.25}, 4 \pi]$\\[4pt]
 & $(m_{Y_{t}} + Y_{t})/f$ & $[e^{-0.5}, 8 \pi]$\\[4pt]
 & $(m_{Y_{b}} + Y_{b})/f$ & $[e^{-8.5}, e^{-0.5}]$\\[4pt]

\midrule

\multirow{7}{*}{LM4DCHM$^{5−5−5}_{5-5}$} & $m_\tau/f$ & $[e^{1.25},4\pi]$ & \multirow{7}{*}{Logarithmic} \\[4pt]
 & $m_{\tilde{\tau}}/f$ & $[e^{1.5},4 \pi]$\\[4pt]
 & $m_{Y\tau}/f$ & $[e^{-8.5},e^{-1.5}]$\\[4pt]
 & $(m_{Y\tau} + Y_\tau)/f$ & $[e^{1.35},8 \pi]$ \\[4pt]
 & $\Delta_{\tau L}/f$ & $[e^{-2.1},e^{-0.5}]$ \\[4pt]
 & $\Delta_{\tau R}/f$ & $[e^{-1.8},e^{-0.2}]$ \\[4pt]

 \midrule

 \multirow{6}{*}{LM4DCHM$^{5−5−5}_{14-10}$} & $m_\tau/f$ & $[e^{-0.5},4\pi]$ & \multirow{6}{*}{Logarithmic} \\[4pt]
 & $m_{\tilde{\tau}}/f$ & $[e^{-0.5},4 \pi]$\\[4pt]
 & $m_{Y\tau}/f$ & $[e^{-1.5},4 \pi]$\\[4pt]
 & $\Delta_{\tau L}/f$ & $[e^{-1.5},4 \pi]$ \\[4pt]
 & $\Delta_{\tau R}/f$ & $[e^{-4.0},e^{-1.5}]$ \\[4pt]

\bottomrule
\end{tabular}
\caption{Parameter ranges and priors used in the scans. The accompanying normalisation factor $f$ is determined after the potential is minimised.}
\label{tab:parameter_bounds}
\end{center}
\end{table}

There are a variety of conditions that limit the possible ranges of our scan parameters:
\begin{itemize}
    \item Mass-dimension parameters (in particular, those of the quark and lepton sectors) cannot be larger than $\Lambda_{f} = 4 \pi f$, the UV cutoff of the effective theory.
    \item The $SO(5)^1$ decay constant $f_1$ must be larger than $f$ by \Cref{decay_constants_relation}, and less than $\sqrt{3} f$ to maintain partial unitarisation of NGB scattering \cite{MarzoccaGeneralCHMs}. The other decay constants $f_{X}$ and $f_{G}$ are constrained by $\frac{f_1}{2} \leq f_{X,G} \leq 2 f_1$ to avoid decoupling any massive gauge bosons.
    \item All gauge couplings are bounded to be between $1$ and $4\pi$ since the composite sector is strongly coupled, and we only perform calculations in the semi-perturbative regime. For $g^{0}_{s}$ to be real we require that $g_G > g_s$.
    \item We also impose the restrictions
    \begin{align}
        \frac{1}{\sqrt{2}} f_{1} g_{\rho} < \Lambda_f, \quad \frac{1}{\sqrt{2}} f_{X} g_{X} < \Lambda_f, \quad \frac{1}{\sqrt{2}} f_{G} g_{G} < \Lambda_f,
    \end{align}
    to avoid vector resonance masses above the cutoff $\Lambda_f$.
\end{itemize}
This leaves some freedom in how we choose the bounds for our fermion scan parameters. Since we scan over these with a logarithmic prior, there is no canonical choice of lower bound. We set the bounds for the quark parameters to be the same as those used in Ref.~\cite{Ethan} in order to facilitate comparison with the results of that study of the M4DCHM. To decide on the bounds for the lepton parameters, we conducted multiple preliminary scans for each model using between $500$ and $1000$ live points, letting the lepton parameters range from an arbitrarily chosen lower bound of $e^{-8.5}$ up to the cutoff scale $4\pi$. With such wide bounds and relatively few live points, these preliminary scans did not have sufficient coverage of the parameter spaces to give consistent convergent results, although they did indicate the more viable regions of parameter space. We used these results to establish narrower bounds for the lepton parameters that restricted the size of the parameter space enough to make comprehensive scans feasible, while still encompassing the viable regions. Our main scans then explored each model within these narrower bounds, listed in Table~\ref{tab:parameter_bounds}, using $4000$ live points. Two scans were performed for each model to verify the results were robust. Our main results come from merging these scans for each model using \texttt{anesthetic} \cite{anesthetic}. The results of each individual scan are presented in \Cref{scan_agreement_appendix} for transparency. The implications of this choice of parameter bounds on our interpretation of the fine-tuning of each model is discussed later in~\Cref{model_statistics_section}.

\subsection{Experimental constraints}
\label{constraints}

We constrain our models using experimental measurements of the SM masses, oblique parameters, Z decay rates, and Higgs signal strengths, exactly as detailed in our previous work \cite{Ethan}, using measurements from Refs.~\cite{PhysRevD.98.030001, Hoang:2014oea, ALEPH:2005ab, Khachatryan:2016vau, Sirunyan:2018koj,ATLAS-CONF-2018-031, CMS:1900lgv}. We do not include flavour constraints because this would require the flavour structure of the theory to be treated in more detail (as was done, for example, in Ref.~\cite{Niehoff:2015iaa}). Since we are presently taking only the third generation to be partially composite due to the complexity of obtaining a convergent global fit over the full parameter space of more detailed models (even with simplifying assumptions about the flavour structure), we leave the study of the full flavour structure to future work.
\begin{itemize}
    \item The Higgs VEV and mass are calculated as detailed in \Cref{Higgs_potential}. Fermion masses are found as the singular values of the fermion mass matrices which are presented in the appendix of \cite{Ethan} and in \Cref{appendix_mass_matrices}. Third generation fermions are identified as the third lightest particles of each type. In addition to the top, bottom and Higgs mass, we also include the $\tau$ mass as a constraint, $M_\tau = 1.77686(12)$~GeV \cite{PhysRevD.98.030001}, to ensure the symmetry breaking and fermion parameters converge onto viable regions.
    \item The Peskin-Takeuchi $S$ and $T$ ``oblique" parameters are important constraints on the EW precision observables of our theories. These restrict the vector boson masses and couplings, as well as limit the potential effects from composite fermions. We assign absolute theoretical uncertainties of $0.05$ and $0.10$ to theoretical predictions of $S$ and $T$ respectively, and assume that these uncertainties are uncorrelated.
    \item The $Z$ boson decay widths
    \begin{align}
        R_b := \frac{\Gamma(Z \rightarrow b \bar{b})}{\Gamma_{\text{had}}}, \qquad R_\ell := \frac{\Gamma_{\text{had}}}{\Gamma(Z \rightarrow \ell \bar{\ell})}, \quad \text{for } \ell = e, \mu, \tau,
    \end{align}
    are well measured and will be modified by the partial compositeness of our third generation fermions. Here
    \begin{align}
    \Gamma_{\text{had}} = \sum\limits_{q = u,d,c,s,b} \Gamma(Z \rightarrow q \bar{q})
    \end{align}
    is the total hadronic width of the $Z$ boson. Given that only third-generation fermions are composite in our models, $R_b$ and $R_\tau$ will be the primary constraints here.
    \item We include measurements for gluon-fusion produced Higgs signal strengths,
    \begin{align}
        \mu^{gg}_X := \frac{\sqbrackets{\sigma (gg \rightarrow h) \text{BR} (h \rightarrow X)}_{\text{exp}}}{\sqbrackets{\sigma (gg \rightarrow h) \text{BR} (h \rightarrow X)}_{|\text{SM}}}.
    \end{align}
    as the ratio of measured Higgs decays into final states, $X=\tau \tau$, $WW$, $ZZ$, and $\gamma \gamma$ to what we would expect in the SM. These measurements serve to constrain the couplings of the Higgs to both the elementary fermions and bosons, including loop contributions from the composite sector. We exclude $\mu^{gg}_{bb}$ from our constraints as the latest results from CMS in this channel are not precise enough to provide meaningful contributions to our analysis \cite{CMS:2018hbb}.
    \item Additionally, we assign steep, one-sided Gaussian likelihoods in order to set lower bounds on the fermion partner masses. Although top and bottom quark partners are now excluded up to ${\sim}1.5$~TeV \cite{ATLAS:2018ziw,CMS:2022fck}, we give them a lower bound of only $500$~GeV in our scans to match the value used in Ref.~\cite{Ethan}. This allows us to compare our results directly with those from the quark-only model of Ref.~\cite{Ethan} and isolate the effects of introducing partially composite leptons. Later, when discussing experimental signatures of our models, we will consider only those points satisfying the current quark partner bounds.

    The bounds we use for the lepton partners are listed in \Cref{tab:lepton_mass_bounds}. Our naming convention for the lepton partners is to use the symbols $N$ (neutral), $L$ (lepton), and $E2$ for partners of electric charge $0$, $\mp 1$, and $\mp 2$ respectively. The lightest $N$ and $L$ partners are denoted $N_4$ and $L_4$, as they are the fourth lightest leptons with their given charge (counting the three SM generations).
\end{itemize}

Further reading on the effects of these constraints on similar models can be found in Refs.~\cite{Banerjee:2017wmg,Banerjee:2020tqc}.

\begin{table}
    \centering
    \begin{tabular}{c|c|l}
        Lepton resonance & Lower Mass Bound & Ref. \\[1.5pt]
        \midrule
        $N$  & $ 90.3 \gev$ & \cite{L3:2001xsz}\\[1.5pt]
        $E2$  & $ 370 \gev$ & \cite{Altmannshofer:2013zba}\\[1.5pt]
        $L_{\mathbf{1}}$ & $ 300 \gev$ & \cite{Bissmann:2020lge}\\[1.5pt]
        $L_{\mathbf{2}}$ & $ 790 \gev$ & \cite{CMS:2019hsm, Falkowski:2013jya, Bissmann:2020lge}\\[1.5pt]
        $L_{\mathbf{3}}$ & $ 225 \gev$ & \cite{Rehman:2020ana}\\[1.5pt]
    \end{tabular}
    \caption{Mass bounds for all lepton partner resonances. $N$, $L$, and $E2$ denote partners of electric charge $0$, $\mp 1$, and $\mp 2$ respectively. Bounds on charge 1 partners are further split according to whether they are $SU(2)_{L}$ singlets, doublets, or triplets $L_{\mathbf{1},\mathbf{2},\mathbf{3}}$.}
    \label{tab:lepton_mass_bounds}
\end{table}

%% file: tex/Results.tex
\section{Results}
\label{results}

In this section, we present the global fits for both models and make comparisons to the quark-exclusive models from our previous work \cite{Ethan}. \Cref{pripos_55_section} and \Cref{pripos_1410_section} showcase plots of the marginalised priors and posteriors for LM4DCHM$^{5-5-5}_{5-5}$ and LM4DCHM$^{5-5-5}_{14-10}$, respectively. In \Cref{model_statistics_section}, we compare and discuss both models with regards to their Bayesian evidences and suggest possible modifications to M4DCHMs for future work. 

\subsection{LM4DCHM$^{5-5-5}_{5-5}$}
\label{pripos_55_section}

\begin{figure}[h]
    \centering
    \includegraphics[width=1\linewidth]{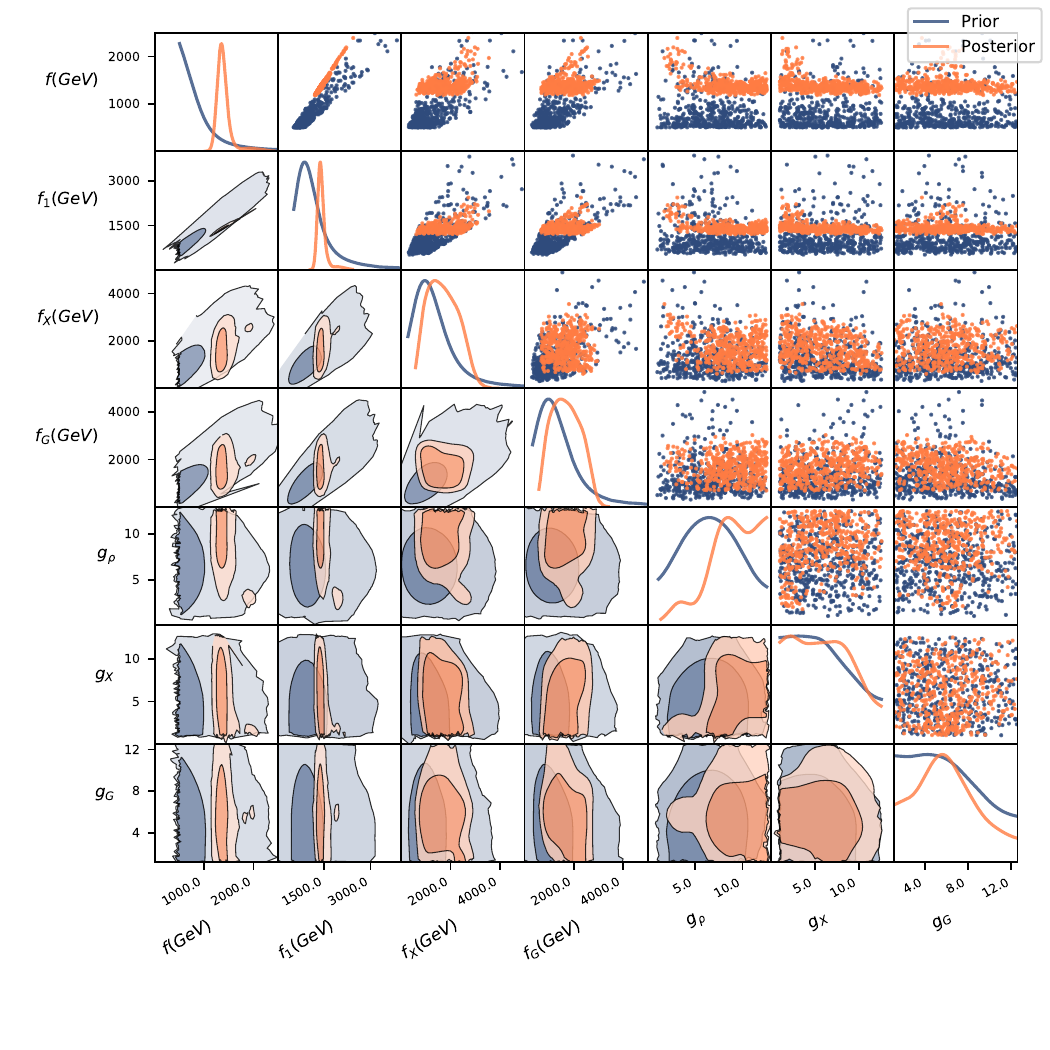}
    \caption{Marginalised priors and posteriors for the gauge sector of the LM4DCHM$^{5-5-5}_{5-5}$.}
    \label{fig:Pripos_55_boson}
\end{figure}

\begin{figure}
    \centering
    \includegraphics[width=1\linewidth]{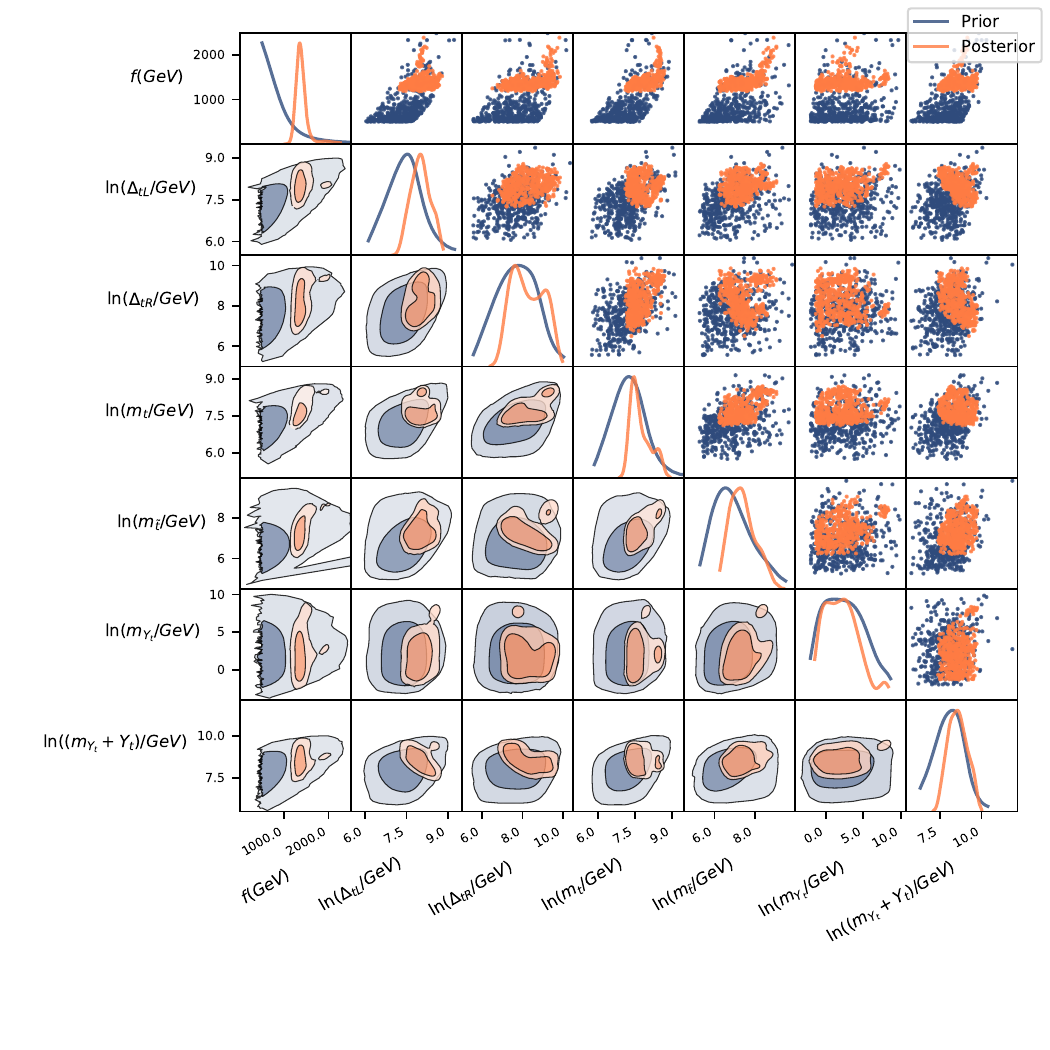}
    \caption{Marginalised priors and posteriors for the top quark sector of the LM4DCHM$^{5-5-5}_{5-5}$.}
    \label{fig:Pripos_55_t}
\end{figure}

\begin{figure}[h]
    \centering
    \includegraphics[width=1\linewidth]{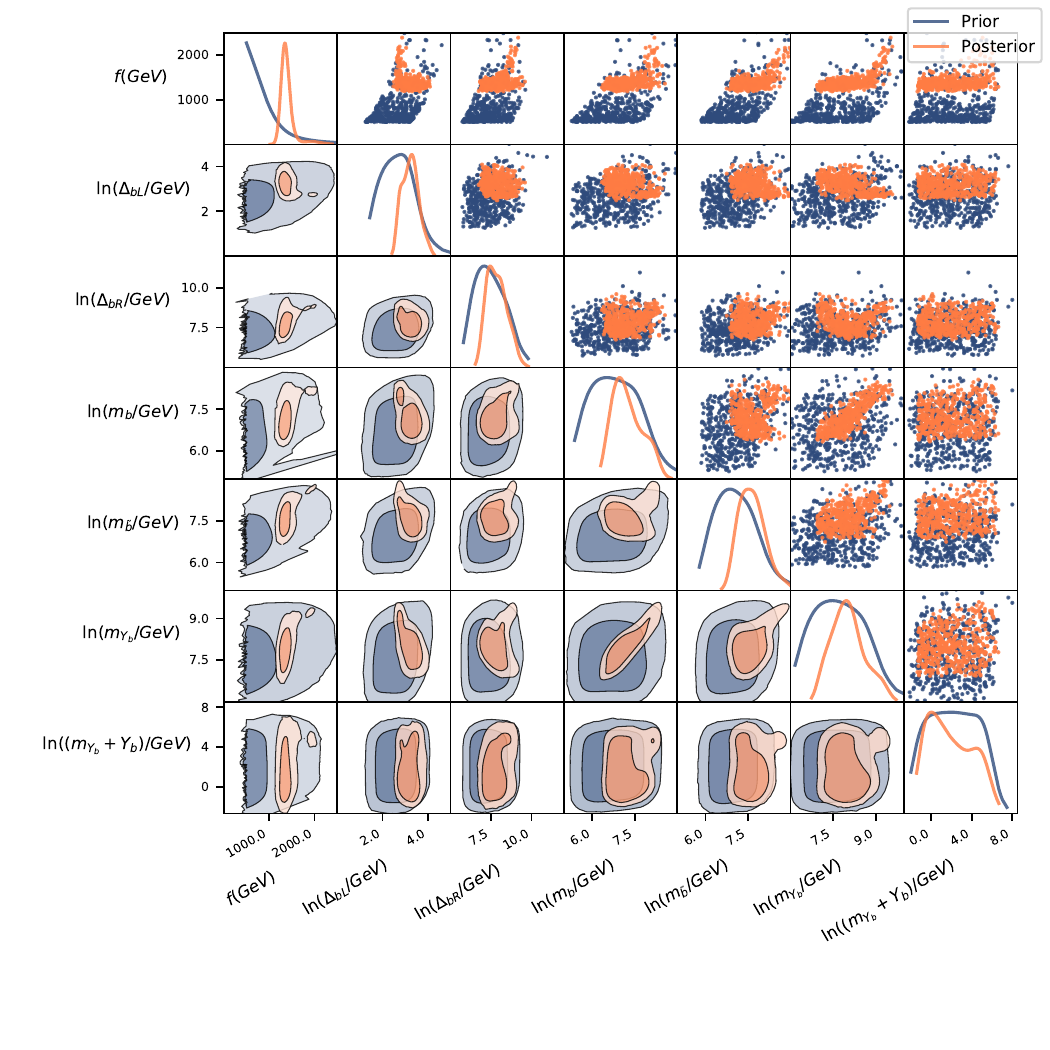}
    \caption{Marginalised priors and posteriors for the bottom quark sector of the LM4DCHM$^{5-5-5}_{5-5}$.}
    \label{fig:Pripos_55_b}
\end{figure}

\begin{figure}[h]
    \centering
    \includegraphics[width=1\linewidth]{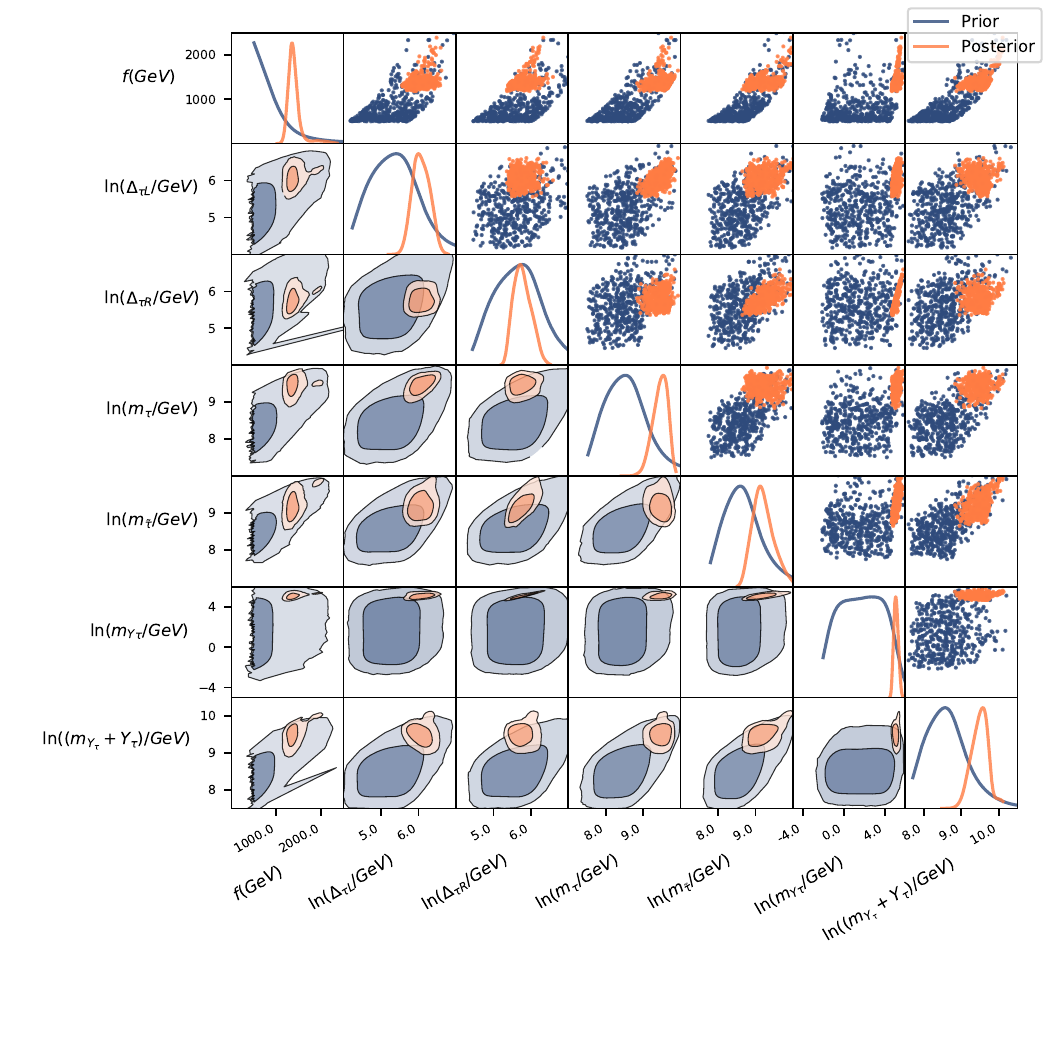}
    \caption{Marginalised priors and posteriors for the tau lepton sector of the LM4DCHM$^{5-5-5}_{5-5}$}
    \label{fig:Pripos_55_tau}
\end{figure}

\Cref{fig:Pripos_55_boson,fig:Pripos_55_t,fig:Pripos_55_b,fig:Pripos_55_tau} depict the 1D and 2D marginalised priors and posteriors for this model. In each figure, the 2D plots above the diagonal show points sampled from the prior and posterior, while those below the diagonal show the 68\% and 95\% credible intervals of the marginalised distributions.

To begin our analysis, we direct our attention to the gauge parameters as presented in \Cref{fig:Pripos_55_boson}. The priors for these parameters favour similar regions as found in the quark-only M4DCHM$^{5-5-5}$ from Ref.~\cite{Ethan} — the decay constants tend to concentrate towards lower values, while gauge couplings span the entire range of $[1,4\pi]$, as expected when only including tau leptons into the mix. Recall that these gauge parameters are sampled uniformly within their imposed ranges; non-uniformities in the priors arise because points that do not generate a symmetry breaking Higgs potential are immediately discarded and do not contribute to the population of live points in the scans. We see the priors for the decay constants strongly favour lower values, as expected by naturalness.

However, there are nuanced differences in the posterior modes. Firstly, there is now a strong correlation between the $SO(5)$ decay constant $f_1$ and $f$, which was not present in the quark-only model. This correlation lies close to the line that marks the lower bound $f_{1} = f$, and is found to be strongly driven by the oblique constraints. This can be understood from \Cref{decay_constants_relation}: the line $f_{1} = f$ corresponds to the $f_{2} \rightarrow \infty$ limit, which decouples the axial vector boson by sending $m_{a} \rightarrow \infty$. The oblique constraints would favour this region because heavier gauge bosons contribute less to the vacuum polarisations of the W and Z bosons. The SM mass constraints (including the newly introduced partially-composite $\tau$ mass constraint) also strongly favour this correlation of $f_{1} \approx f$, showing why there may be a difference with the quark-only model. The reason for this latter correlation is not self-evident and to understand it would require extensive additional investigation. This is left for future work.

We should perhaps stress that it is difficult to pinpoint the physical reasons why certain parameter ranges are favoured by each constraint. The vast majority of observables in these models are incredibly non-linear functions of the Lagrangian parameters, and can only be calculated by numerical minimisation of the Higgs potential and subsequent diagonalisation of the large mass matrices. In certain limited cases there may be heuristic explanations for specific behaviour, and we will try to provide these where possible, as we did above, but in most cases we can only resort to reporting \textit{what} constraints are responsible for the posteriors, and not \textit{why} they are responsible.

Looking at $f_{1}$ further, it exhibits a higher level of fine-tuning than in the quark-only model, with its posterior distribution focused within a narrow region of 1.35 TeV to 1.5 TeV, with regions beyond $\sim$1.6 TeV now largely excluded. This behaviour is found to be mainly driven by the $Z \rightarrow \tau \bar{\tau}$ decay constraint, $R_\tau$, although it should be noted that all SM masses also favour 1.25 TeV $\lesssim f_1 \lesssim$ 1.75 TeV, similar to the quark-only scans. 

Also contrasting the quark only model, posteriors of the SO$(5)^1$ coupling $g_\rho$ now largely cover only values $\gtrsim 6$ rather than the entire prior. This is found to be due to SM mass constraints disfavouring values $\lesssim 6$, despite no obvious correlation between the two. Posteriors for the $U(1)_X$ and $SU(3)_C$ decay constants and couplings ($f_X, g_X, f_G, g_G$) generally agree with the quark-only scans, showing no clear preference within their imposed prior regions.

In regards to the fine-tuning and performance of this model, we will additionally discuss a smaller mode of the posterior volume that (as we'll see in \Cref{exp_sig_section}) contains points much more favoured by direct collider constraints, even though these weren't included as constraints in the scans. This mode contributed substantially to the difficulty of obtaining consistent results, only being found in half of our scans\footnote{Results from individual scans are shown in \Cref{scan_agreement_appendix}.}. This can be seen in \Cref{fig:Pripos_55_boson} at e.g. $f \approx 2$ TeV and will be useful to our discussion in \Cref{model_statistics_section} and \Cref{exp_sig:higgs_sig_strengths}. In general, the $Z$ decay constraints $R_{b, e, \mu, \tau}$ and SM masses show preference around $f \sim 2$ TeV, noting that the latter still favours the main posterior peaks more than this smaller mode. 

The posteriors for the top quark sector parameters in \Cref{fig:Pripos_55_t} are mostly contained within their priors' $68$\% peaks, indicating that there is no significant constraint favouring the outside of this region. The plots are broadly in agreement with those from the quark-only models, with the left-handed composite coupling $\Delta_{tL}$ and the mass parameters $m_t$, $m_{\tilde{t}}$ peaking between approximately 1 to 8 TeV, and the right-handed composite coupling $\Delta_{tR}$ having the same lower bound but also showing preference for large values up to $\sim$17~TeV. All said posteriors are mainly influenced by SM mass measurements and oblique constraints, with two of the four Z decay constraints $R_\tau$ and $R_b$ slightly contributing to their preferred posterior regions. The bottom quark sector parameters, depicted in \Cref{fig:Pripos_55_b}, also do not show much difference from the quark-only model.

The tau sector is depicted in \Cref{fig:Pripos_55_tau}. There is a clear pattern here: most of the parameters, save for the right-handed tau-composite coupling $\Delta_{\tau R}$, show an affinity for a parameter region outside of their prior peaks, indicating a high degree of fine-tuning. The $\tau$ posteriors also span a smaller range than the boson and quark parameters, especially $m_{Y \tau}$. This behaviour is largely driven by oblique constraints, SM mass measurements, as well as $R_\tau$, and to a lesser degree the other Z decay constraints $R_b, R_e,$ and $R_\mu$. Compared to the top and bottom quarks, the $\tau$ is less strongly coupled to the composite sector as measured by the ratio between its link and mass parameters, $\Delta / m$, which is to be expected from its smaller mass. 

\clearpage
\subsection{LM4DCHM$^{5-5-5}_{14-10}$}
\label{pripos_1410_section}

\begin{figure}[h]
    \centering
    \includegraphics[width=1\linewidth]{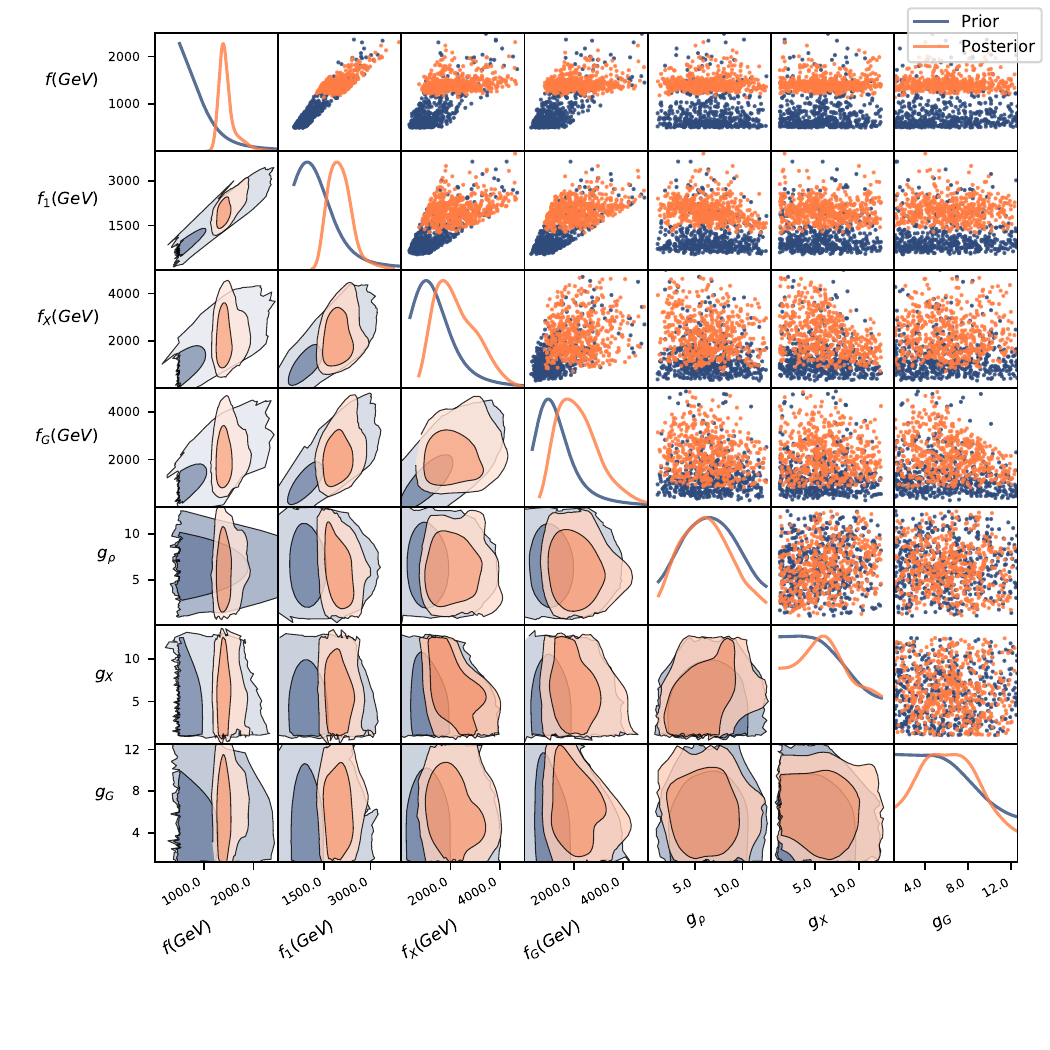}
    \caption{Marginalised priors and posteriors for the boson sector of the LM4DCHM$^{5-5-5}_{14-10}$.}
    \label{fig:Pripos_1410_boson}
\end{figure}

\begin{figure}[h]
    \centering
    \includegraphics[width=1\linewidth]{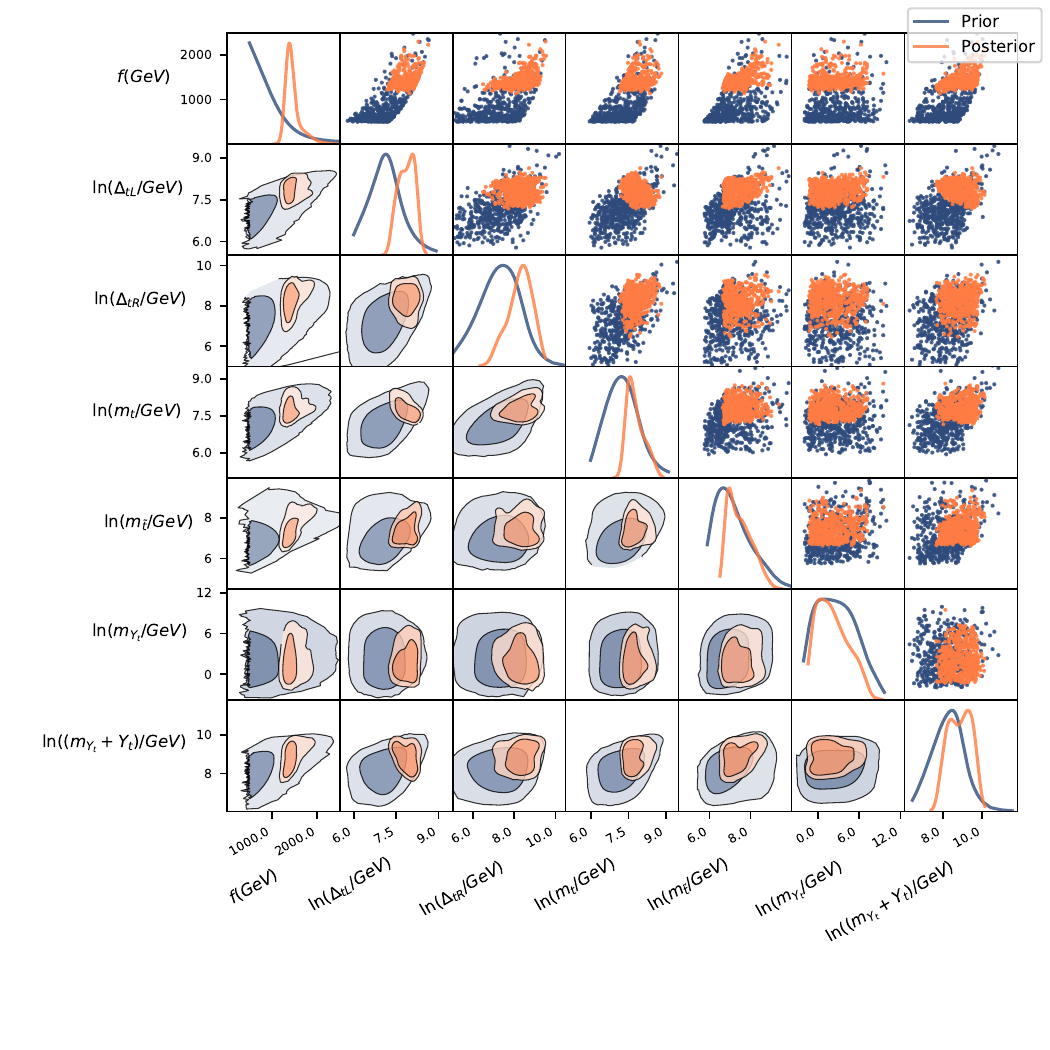}
    \caption{Marginalised priors and posteriors for the top quark of the LM4DCHM$^{5-5-5}_{14-10}$.}
    \label{fig:Pripos_1410_t}
\end{figure}

\begin{figure}[h]
    \centering
    \includegraphics[width=1\linewidth]{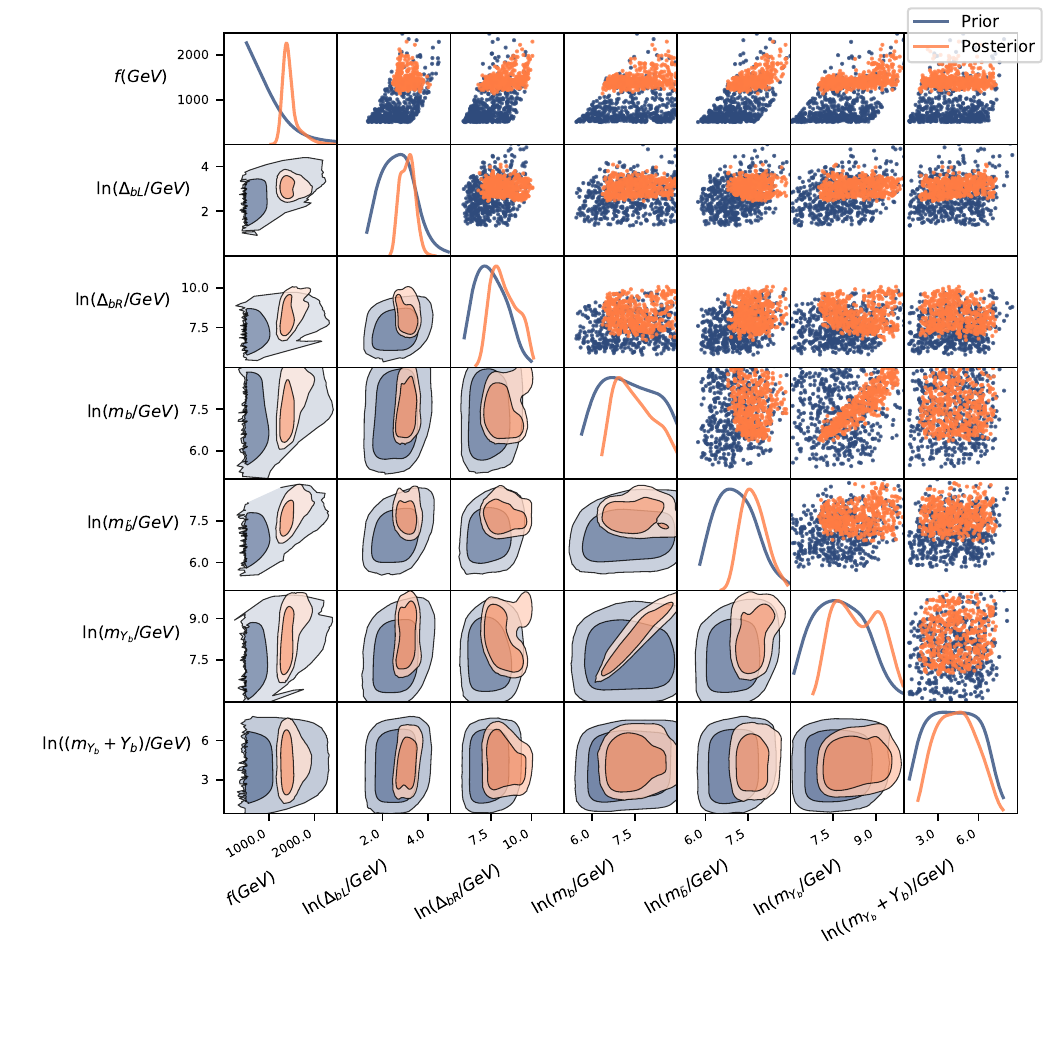}
    \caption{Marginalised priors and posteriors for the bottom quark of the LM4DCHM$^{5-5-5}_{14-10}$.}
    \label{fig:Pripos_1410_b}
\end{figure}

\begin{figure}[h]
    \centering
    \includegraphics[width=1\linewidth]{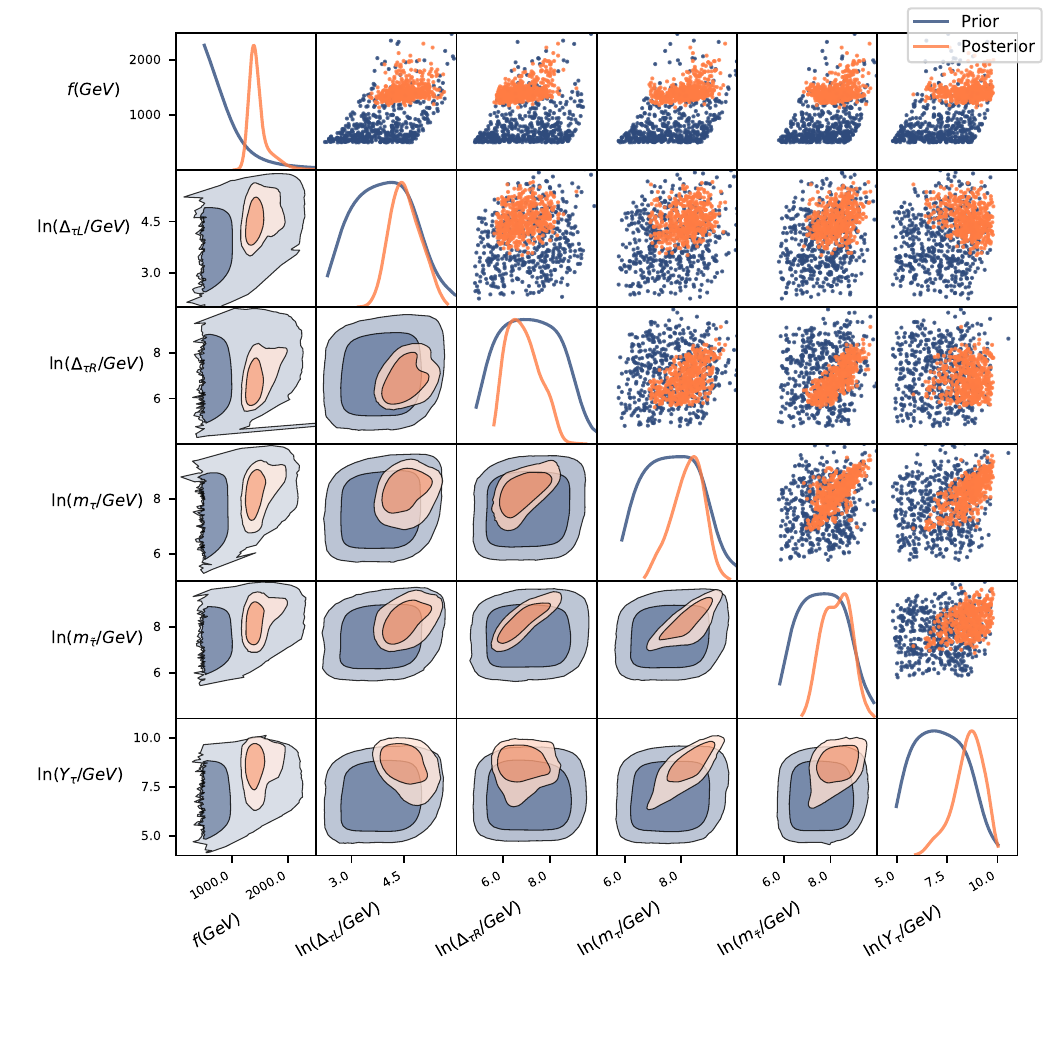}
    \caption{Marginalised priors and posteriors for the tau lepton of the LM4DCHM$^{5-5-5}_{14-10}$.}
    \label{fig:Pripos_1410_tau}
\end{figure}

\Cref{fig:Pripos_1410_boson,fig:Pripos_1410_t,fig:Pripos_1410_b,fig:Pripos_1410_boson,fig:Pripos_1410_tau} show our results for the LM4DCHM$^{5-5-5}_{14-10}$. We start with the boson sector, shown in \Cref{fig:Pripos_1410_boson}. Similarly to the previous model, the posteriors of the decay constants $f$ and $f_1$ peak at higher values than their priors. The posterior for $f$ is mostly localised around $1.5$~TeV, while $f_1$ is significantly less fine-tuned than in the previous model, with its posterior favouring regions between $1.4$ and $2.75$~TeV. This largely aligns with the quark-only scans. From a theoretical perspective, smaller decay constants are more natural, but are disfavoured by EW precision tests.

\Cref{fig:Pripos_1410_t} shows the top quark sector. As expected, the marginalised posteriors are not significantly different to those in the previous model (\Cref{fig:Pripos_55_t}) where the SM masses and oblique constraints are again the primary influencers for these posterior regions.


This behaviour repeats again for the bottom quark sector in \Cref{fig:Pripos_1410_b}, having results very similar to \Cref{fig:Pripos_55_b}. There is a slight difference in the posterior for $\Delta_{b L}$, which here excludes regions from $47 \gev$ to $55\gev$, and also $m_{Y b}$, which now has an extended posterior mode beyond the prior up to ${\sim}22$ TeV. The oblique and SM mass constraints are again the factors affecting this.

Lepton sector results are shown in \Cref{fig:Pripos_1410_tau}. We see strong correlations between the parameter pairs ($m_\tau$, $m_{\tilde{\tau}}$), ($m_\tau$, $Y_\tau$), and ($m_{\tilde{\tau}}$, $\Delta_{\tau R}$), and a weaker correlation between ($\Delta_{\tau R}$, $m_\tau$). The lower bounds of these parameters are mainly driven by the $\tau$-lepton mass constraint, and to a lesser extent the bottom quark mass. In particular, both the on-diagonal mass parameters $m_\tau$ and $m_{\tilde{\tau}}$ disfavour regions $\lesssim 1$~TeV as this would yield $\tau$-lepton masses that are too light.

The lepton posterior regions are as follows:

\begin{align*}
\begin{array}{rcl}
    40 \gev \lesssim & \Delta_{\tau L} & \lesssim 400 \gev,\\[0.1cm]
    270 \gev \lesssim & \Delta_{\tau R} & \lesssim 4.8 \tev,\\[0.1cm]
    1.1 \tev \lesssim & m_\tau & \lesssim 36.3 \tev,\\[0.1cm]
    900 \gev \lesssim & m_{\tilde{\tau}} & \lesssim 12.5 \tev,\\[0.1cm]
    750 \gev \lesssim & Y_\tau & \lesssim 22 \tev.
\end{array}
\end{align*}

These posteriors tend to favour higher values, although without needing as much fine-tuning as in the previous model, since they span a much bigger portion of the prior volume despite the priors in this model having been defined to be larger (on a logarithmic scale). The main constraints causing this behaviour are the $\tau$-lepton mass, oblique constraints, $\mu^{gg}_{\tau \tau}$, and the Z boson decay fraction into $\tau \tau$ pairs, $R_\tau$. All these are understandable since $m_\tau$ will receive contributions from all the aforementioned parameters (see \Cref{appendix_mass_matrices} mass matrices). 
This model also predicts a much smaller compositeness scale for the $\tau$ compared to the top and bottom quarks, especially so for the left-handed component based on $\Delta_{\tau_L}$ and $m_\tau$. In both models, the top quark has a larger compositeness than the bottom as is to be expected.

To summarise, the introduction of the tau leptons not only introduces exclusive constraints, but also affects the fine-tuning of the boson sector and, to a lesser extent the quark sector. The LM4DCHM$^{5-5-5}_{5-5}$ emerges as more fine-tuned than the LM4DCHM$^{5-5-5}_{14-10}$, with not only increased specificity (in terms of parameter regions) in bosonic parameters such as $f_1$, but also across entire sectors as with the $\tau$ lepton parameters, despite the smaller prior volume.

\subsection{Discussion and model statistics}
\label{model_statistics_section}

\begin{table}[h]
\begin{center}
\begin{tabular}{l|ccccc}
Model & $\ln(\mathcal{Z}$)  & $ \langle \ln(\mathcal{L}) \rangle_{P}$ & $\max \ln(\mathcal{L})$ & $D_{KL}$ \\[2pt]
\midrule
LM4DCHM$^{5-5-5}_{5-5}$   & $-45.60 \pm 0.06$  & -17.27  & -10.79 & 28.33 \\[2pt]
LM4DCHM$^{5-5-5}_{14-10}$ & $-36.30 \pm 0.05$  & -14.63  & -9.13  & 21.67
\end{tabular}
\end{center}
\caption{Statistics from the combined Bayesian scans of each model, using the priors from \Cref{tab:parameter_bounds}.}
\label{tab:Bayesian_statistics}
\end{table}

We now turn our attention to the Bayesian statistics calculated from our scans in order to compare the two models, both in terms of Bayesian evidence and fine-tuning. The statistics are presented in \Cref{tab:Bayesian_statistics}, including the log-evidence $\ln(\mathcal{Z})$, the posterior-averaged log-likelihood $\langle \ln(\mathcal{L}) \rangle_{P}$, the maximum log-likelihood found in the scans $\max \ln(\mathcal{L})$, and the Kullback-Leibler (KL) divergence $D_{KL}$ for each model, defined in \Cref{scanning_method}.

Keeping in mind that both models are capable of fitting the experimental data for appropriate parameter choices at 3$\sigma$ from a frequentist perspective,
the Bayesian
evidence of the LM4DCHM$^{5-5-5}_{14-10}$ is four orders of magnitude greater than that of the LM4DCHM$^{5-5-5}_{5-5}$, indicating a decisive preference for the LM4DCHM$^{5-5-5}_{14-10}$ from a Bayesian perspective. Recall that the Bayesian evidence is a comprehensive measure that balances a model's fit to experimental data with its theoretical naturalness. A larger evidence could be due to either a higher posterior-averaged log-likelihood or to a lower KL divergence (i.e., fine-tuning), since these are related by \Cref{Eq:lnZ}. In our study both of these factors contribute to the higher evidence of the LM4DCHM$^{5-5-5}_{14-10}$, as can be seen from \Cref{tab:Bayesian_statistics}. Its higher average log-likelihood indicates that the model can fit experimental constraints better over its posterior distribution, and its lower KL divergence indicates that its posterior covers a larger region of its prior volume, as was seen above. This fits with the expectation that the LM4DCHM$^{5-5-5}_{14-10}$ is less finely-tuned due to its effectiveness in generating electroweak symmetry breaking from the fermionic contributions to the Higgs potential per \Cref{eq:Higgs_potentialFF}. Note that for both models, the fine-tuning is the main factor contributing to the evidence, as opposed to the posterior-averaged log-likelihood. Indeed, the average log-likelihoods and especially the maximum log-likelihoods of the two models are quite similar as seen in Table \ref{tab:Bayesian_statistics}, indicating that they provide more or less equally good fits to the data and so are on comparable footing from a frequentist viewpoint. In other words both models have parameter sets that fit the experimental data.

\begin{table}
\begin{center}
\begin{tabular}{l|cccc}
Model & $\ln(\mathcal{Z})$  & $ \langle \ln(\mathcal{L}) \rangle_{P}$ & $\max \ln(\mathcal{L})$ &  $D_{KL}$ \\[2pt]
\midrule
LM4DCHM$^{5-5-5}_{5-5}$   & $-65.06$ & -16.75 & -10.79 & 48.31 \\[2pt]
LM4DCHM$^{5-5-5}_{14-10}$ & $-50.34$ & -15.37 & -9.13  & 34.97
\end{tabular}
\end{center}
\caption{Statistics from the combined Bayesian scans of each model, with the samples re-weighted as if all parameters had been given uniform priors with the same bounds as in \Cref{tab:parameter_bounds}.}
\label{tab:Bayesian_statistics_uniform}
\end{table}

While imposing logarithmic priors on the fermion parameters allowed us to efficiently explore scales across many orders of magnitude, this was ultimately an arbitrary choice. It is important to see how the above conclusions depend on this choice. Fortunately, the sampling method in our scans allows us to post-process our samples as if they had been taken from different prior distributions. The model statistics that would be obtained if the fermion parameters were to have been sampled from a uniform prior, for example, are shown in \Cref{tab:Bayesian_statistics_uniform}. The LM4DCHM$^{5-5-5}_{14-10}$ now has an evidence six orders of magnitude larger than the LM4DCHM$^{5-5-5}_{5-5}$, indicating it is the preferred model to an even higher degree under these conditions. This great disparity in evidences is due almost entirely to the much larger fine-tuning of the LM4DCHM$^{5-5-5}_{5-5}$; it is only slightly disfavoured on the basis of average log-likelihood. Hence, the main conclusion that the LM4DCHM$^{5-5-5}_{14-10}$ is decisively preferred is unchanged, although the exact degree and reason for superiority seems to be strongly prior dependent.

\begin{table}
\begin{center}
\begin{tabular}{l|ccccc}
Model & $\ln(\mathcal{Z}$)  & $ \langle \ln(\mathcal{L}) \rangle_{P}$ & $\max \ln(\mathcal{L})$ & $D_{KL}$ \\[2pt]
\midrule
LM4DCHM$^{5-5-5}_{5-5}$   & $-67.65 \pm 0.20$  & -26.14  & -12.08 & 41.51 \\[2pt]
LM4DCHM$^{5-5-5}_{14-10}$ & $-41.93 \pm 0.16$  & -15.13  & -9.13  & 26.80
\end{tabular}
\end{center}
\caption{Statistics for each model, using much wider bounds $[e^{-8.5},4\pi]$ or $[e^{-8.5},8\pi]$ and logarithmic prior distributions for all fermion parameters.}
\label{tab:Bayesian_statistics_wide_bounds}
\end{table}

It is also important to note that constraining our priors as in \Cref{tab:parameter_bounds} has led to an artificial enhancement of the Bayesian evidences, since we deliberately chose these bounds to focus on regions with better likelihoods as was explained in \Cref{scan_parameters}. To quantify the prior bound dependence of the results, we have performed some small-scale scans of both models using much wider prior bounds on all fermion parameters, ranging from an arbitrarily chosen lower bound of $e^{-8.5}$ to the maximum theoretical cutoff of $4\pi$ (or $8\pi$ for combined parameters such as $m_{Y_{\tau}}+Y_{\tau}$). The resulting parameter spaces are larger (in log space) than those in the main scans by a factor of $3.8 \times 10^{11}$ for the LM4DCHM$^{5-5-5}_{5-5}$, and $2.9 \times 10^{9}$ for the LM4DCHM$^{5-5-5}_{14-10}$. With such vast parameter spaces we were not able to obtain consistent convergent results, but these scans were sufficient to estimate the evidences given these bounds. As such, we regard these as only supplementary to our main results. The statistics from these scans are listed in \Cref{tab:Bayesian_statistics_wide_bounds}.

As expected, the evidence for each model suffers greatly from this expansion of the parameter space. Since the evidence is simply the prior-averaged likelihood, this confirms that our main scans were given unfairly advantageous prior bounds within the wider available spaces. However, the regions of higher average likelihood identified in our main scan have a completely negligible effect in these wider scans - they turn out to be responsible for only $0.2$\% and $10^{-5}$\% of the total evidence of the LM4DCHM$^{5-5-5}_{5-5}$ and LM4DCHM$^{5-5-5}_{14-10}$ in \Cref{tab:Bayesian_statistics_wide_bounds} respectively. Hence, even when given equal prior bounds, which significantly reduces the possibility that one model has been artificially constrained to a much more favourable region than the other, the LM4DCHM$^{5-5-5}_{14-10}$ is once again the clearly preferred model from the standpoint of both posterior-averaged log-likelihood and fine-tuning. 

As mentioned in the previous section, there was a separate posterior mode of $f$ in the LM4DCHM$^{5-5-5}_{5-5}$ that could only be reproduced in half of our scans. Visible in \Cref{fig:Pripos_55_boson} which entails a larger $f$ decay constant at around $2 \tev$, it predicted favourable phenomenology for this model and is responsible for the vast majority of points within 3$\sigma$ of the experimental constraints found in the scans, which we discuss further in \Cref{exp_sig:higgs_sig_strengths}. However, these modes cover only a small portion of the model's parameter volume, and its resulting lower evidence is consistent with the increased difficulty in obtaining robust convergent results.

%% file: tex/Pheno.tex
\section{Experimental signatures}
\label{exp_sig_section}
Here we examine the predicted phenomenology for ``valid" points in each model — points in our scans that have simultaneously satisfied all constraints and collider bounds at the $3\sigma$ level. We further restrict to points that additionally satisfy recent quark resonance bounds of 1.54 and 1.56 TeV for the top and bottom partners respectively \cite{ATLAS:2018ziw,CMS:2022fck}, which were not implemented in the scans as explained in \Cref{constraints}. The scans have found 843 valid points within the LM4DCHM$^{5-5-5}_{5-5}$, and 976 valid points in the LM4DCHM$^{5-5-5}_{14-10}$. Details regarding individual scans contributing to our results are given in \Cref{scan_agreement_appendix}.

In \Cref{cross_sections} we analyse the predicted production cross sections of BSM particles for these valid points for both models to anticipate their potential signals in future collider experiments. Additionally, we assess their predicted gluon-fusion produced Higgs signal strengths to see how the new physics introduced in these models modifies these decay rates compared to the SM in \Cref{exp_sig:higgs_sig_strengths}.

Of the two 4000-point LM4DCHM$^{5-5-5}_{5-5}$ scans, the one with the extra posterior mode as mentioned in \Cref{results} accounted for almost 99\% of valid points. As a reminder, this smaller mode lies between 1.75 TeV $\lesssim f \lesssim 2.5$ TeV and appears due to more favourable predictions for Z decay ratios, while points with larger $f$ would fail SM mass constraints. We further discuss this in \Cref{exp_sig:higgs_sig_strengths}.

\subsection{Composite resonances}
\label{cross_sections}

Both of the models we are considering contain fields beyond the SM, including composite gauge bosons and heavy quark and lepton partners. Here we will only give details of the heavy lepton decay signatures, as the quark partner and gauge boson signatures are almost entirely the same as in our previous work Ref.~\cite{Ethan}. The only point to note is that the leptonic LM4DCHM$^{5-5-5}_{5-5}$ model now predicts an upper mass bound of $\sim$5~TeV for the lightest neutral composite vector boson $Z_3$ and obviously quark resonances that satisfy the bounds from Refs.~\cite{ATLAS:2018ziw,CMS:2022fck} as stated earlier in this section.

The lightest $N$ and $L$ partners are denoted $N_4$ and $L_4$, as they are the fourth lightest leptons with their given charge (counting the three SM generations).

\subsubsection*{Decay channels}

\begin{figure}[h]
\centering
\begin{subfigure}{.49\linewidth}
  \centering
  \includegraphics[width=1\textwidth]{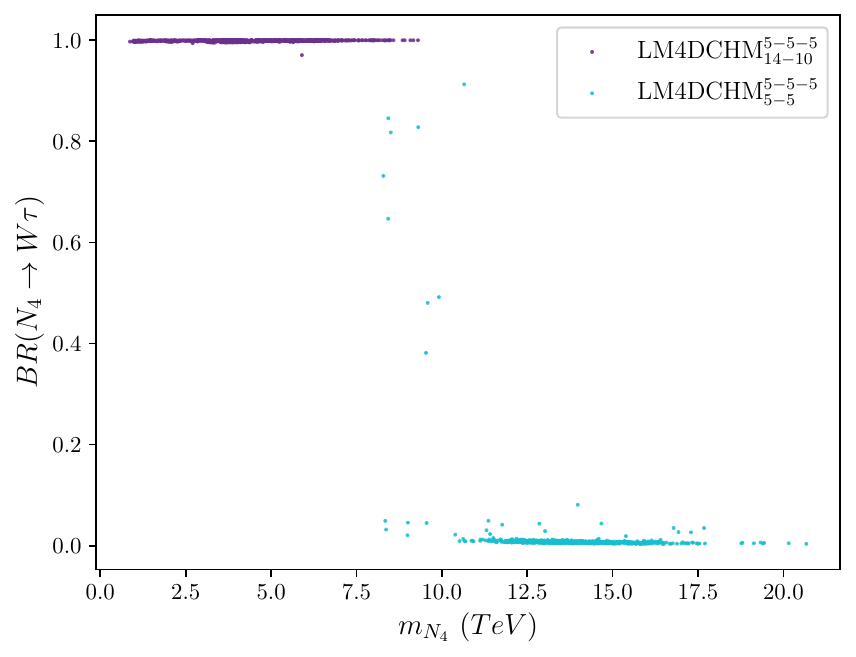}
\end{subfigure}
\begin{subfigure}{.49\linewidth}
  \centering
  \includegraphics[width=1\textwidth]{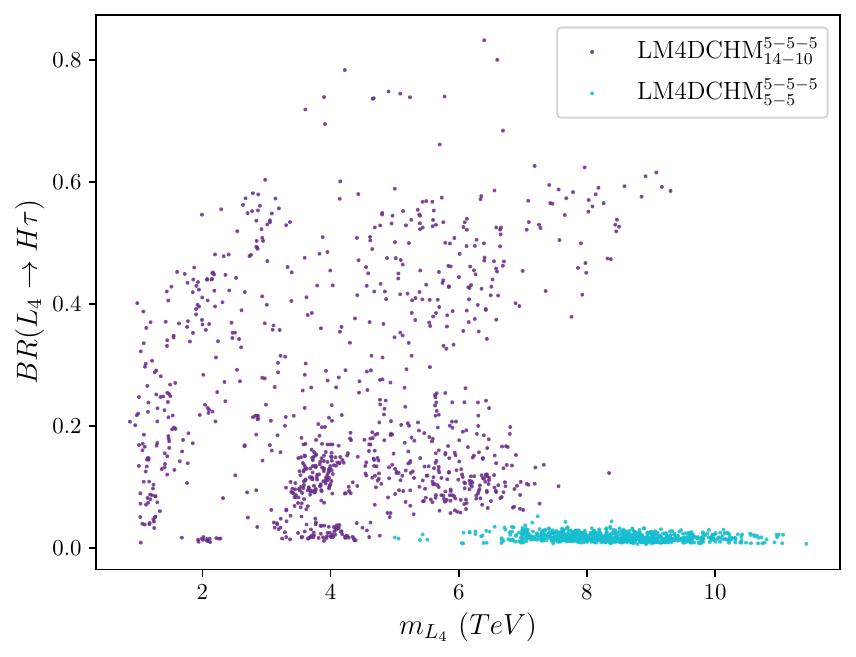}
\end{subfigure}  
\begin{subfigure}{.49\linewidth}
  \centering
  \includegraphics[width=1\textwidth]{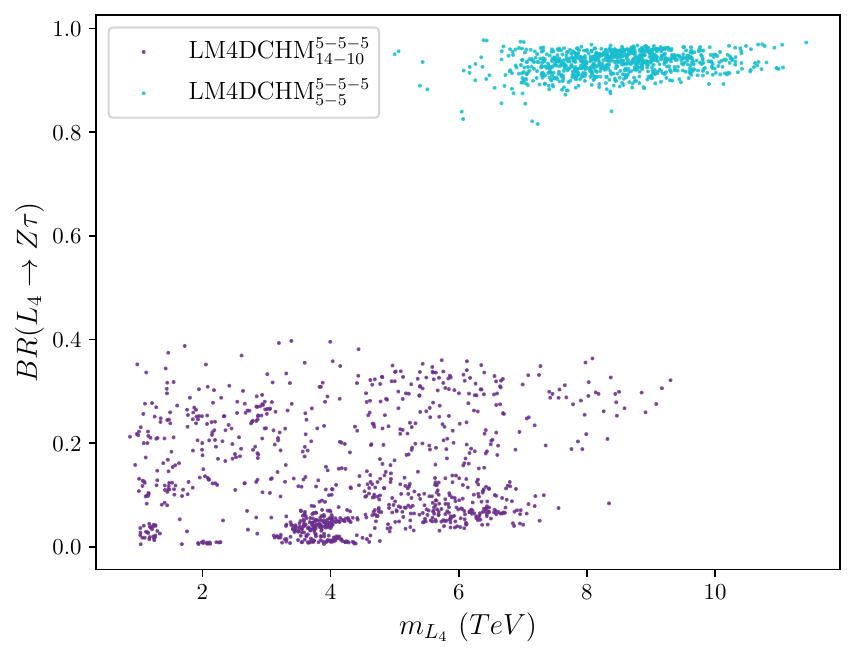}
\end{subfigure}
\begin{subfigure}{.49\linewidth}
  \centering
  \includegraphics[width=1\textwidth]{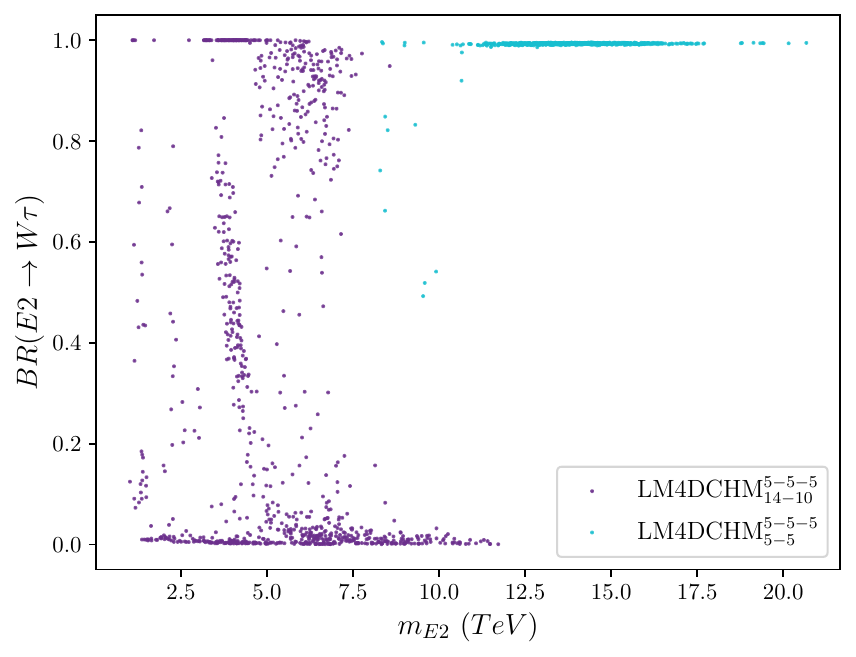}
\end{subfigure}
\caption{Branching ratios for the decays of the lepton resonances $N_4$, $L_4$ and $E2$ into SM-pair final states for valid points of both models.}
\label{fig:lepton_BR}
\end{figure}

Branching ratios for the various decay channels of the heavy lepton partners are shown in \Cref{fig:lepton_BR}, while estimates of the production cross sections of these partners at the $13$~TeV LHC are shown in \Cref{fig:N_4_xsec,fig:L_4_xsec,fig:E2_xsec}. The green shaded regions of \Cref{fig:N_4_xsec,fig:L_4_xsec,fig:E2_xsec} highlight those points that are potentially discoverable with an integrated luminosity of 3000~fb$^{-1}$ of LHC proton-proton collision data that will be available after the High-Luminosity (HL) LHC has completed its full run.
 
In the LM4DCHM$^{5-5-5}_{5-5}$, $N_4$ particles mostly fall within the 8 to 17.5~TeV range, though outliers are present up to 22~TeV. These particles share nearly identical masses with the $E2$ particles of the same model, since these masses become degenerate in the limit $\Delta_{\tau L} \ll m_\tau, m_{Y_\tau}, m_{\tilde{\tau}}$ as can be seen from the mass matrices in \Cref{appendix_mass_matrices}. Those with masses less than ${\sim}13.5$~TeV are likely to decay to a $W \tau$ pair. In contrast, the LM4DCHM$^{5-5-5}_{14-10}$ predicts the $N_4$ mass to be anywhere between 1~TeV and ${\sim}9$~TeV, with most points having a mass that lies between 2 to 6~TeV, with a decay that is dominated by the $W \tau$ channel.

$L_4$ is typically the lightest composite state across both models. In the LM4DCHM$^{5-5-5}_{5-5}$ model, $L_4$ masses typically range from 5 to ${\sim}11$~TeV and the particle largely decays via the $Z \tau$ channel. In the LM4DCHM$^{5-5-5}_{14-10}$, $L_4$ states span from 1~TeV to ${\sim}9$~TeV and tend to be more likely to decay via the $H \tau$ channel (with a branching ratio $\lesssim 0.6$), but with a significant fraction decaying into a $Z \tau$ pair (branching ratio $\lesssim 0.4$). There is no clear relation between these branching ratios and the associated mass.

Doubly-charged particles $E2$ are predicted to almost always decay into a $W \tau$ pair for masses $\gtrsim 11$~TeV in the LM4DCHM$^{5-5-5}_{5-5}$, though for lower masses they also decay significantly into $W L_4$. The $E2$ decay patterns are less straightforward in the LM4DCHM$^{5-5-5}_{14-10}$, where it decays appreciably often into $W \tau$ only for a fraction of the points with masses below $8$~TeV, but otherwise almost never decays through this channel, instead tending to decay into other composite states such as $W L_{4}$ and $W_{2} \tau$, which then decay further into SM states.

\subsubsection*{Cross sections}

\begin{figure}
    \centering
    \includegraphics[width=0.75\textwidth]{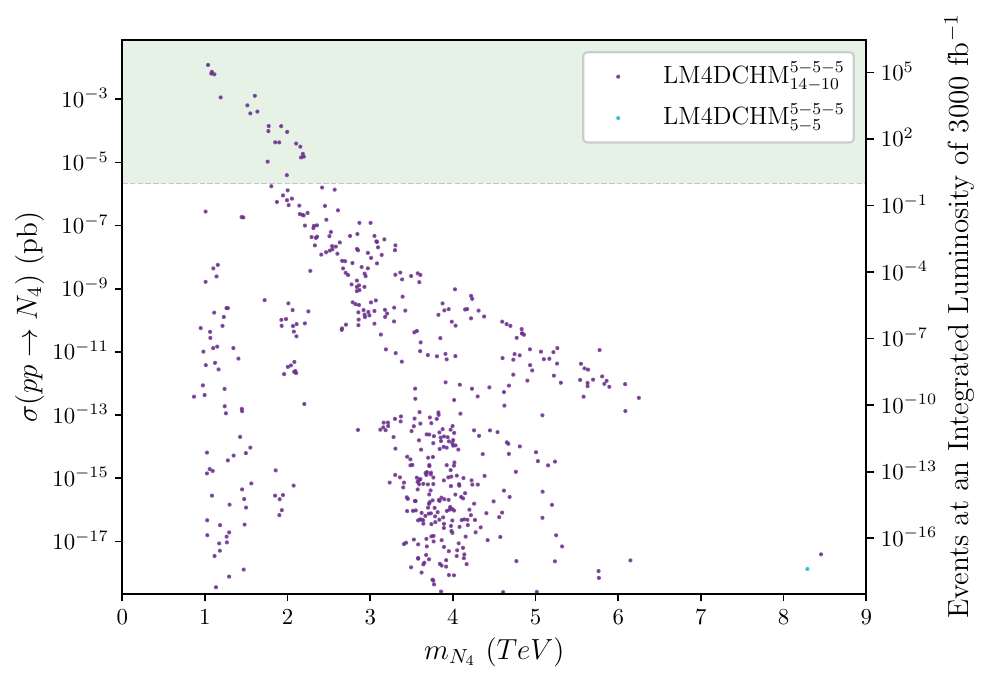}
    \caption{Cross sections for the process $pp \rightarrow W/Z \rightarrow N_4$ at $\sqrt{s} = 13$ TeV for valid points in both models. The axis on the right shows the expected number of $N_4$ particles produced at an integrated luminosity of 3000 fb$^{-1}$. The green shaded region is that where the expected number of particles is $\geq 1$.}
    \label{fig:N_4_xsec}
\end{figure}

\begin{figure}
    \centering
    \includegraphics[width=0.75\textwidth]{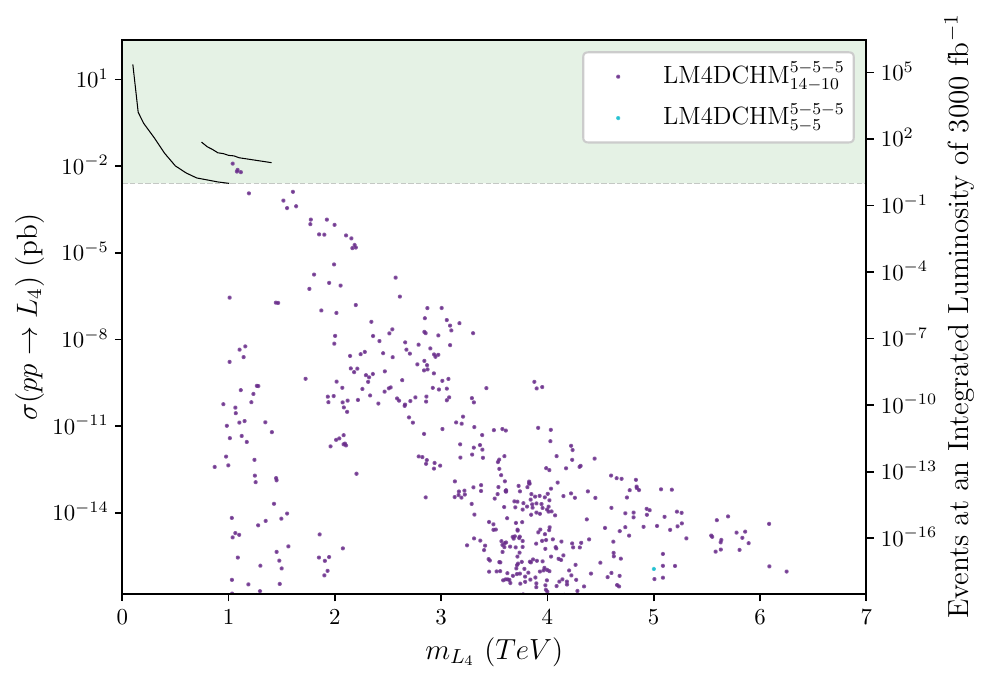}
    \caption{Cross sections for the process $pp \rightarrow W/Z \rightarrow L_4$ at $\sqrt{s} = 13$ TeV for valid points in both models. The axis on the right shows the expected number of $L_4$ particles produced at an integrated luminosity of 3000 fb$^{-1}$. The green shaded region is that where the expected number of particles is $\geq 1$. Black lines indicate bounds placed by collider search constraints.}
    \label{fig:L_4_xsec}
\end{figure}

\begin{figure}
    \centering
    \includegraphics[width=0.75\textwidth]{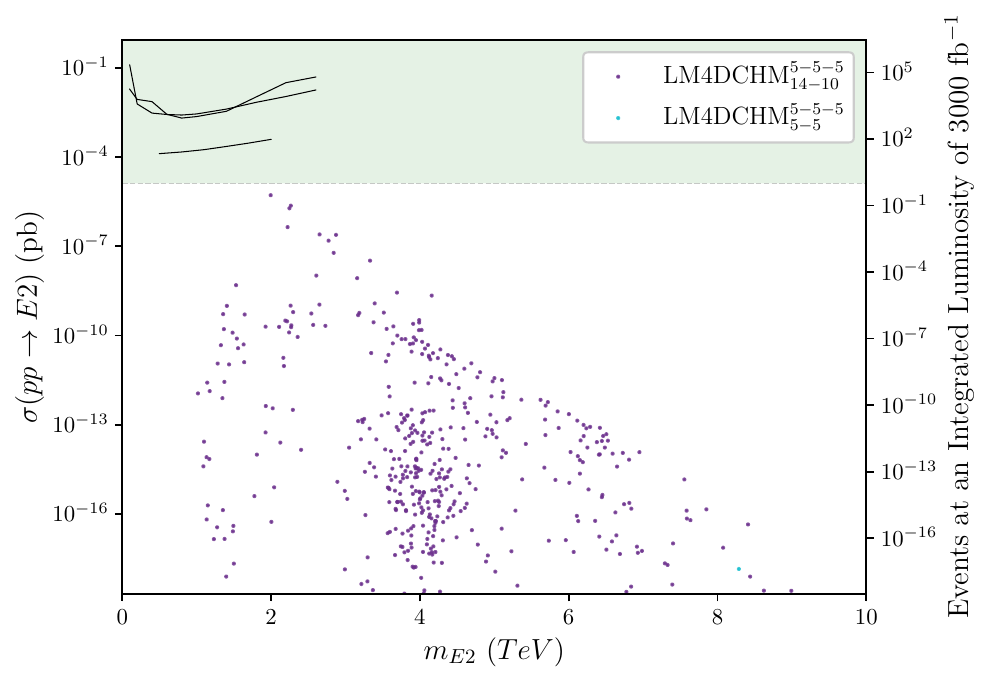}
    \caption{Cross sections for the process $pp \rightarrow W/Z \rightarrow E2$ at $\sqrt{s} = 13$ TeV for valid points in both models. The axis on the right shows the expected number of $E2$ particles produced at an integrated luminosity of 3000 fb$^{-1}$. The green shaded region is that where the expected number of composite Higgs production events is $\geq 1$. Black lines indicate bounds placed by collider search constraints.}
    \label{fig:E2_xsec}
\end{figure}

We should stress that the cross sections of \Cref{fig:N_4_xsec,fig:L_4_xsec,fig:E2_xsec} are only approximate, and certainly underestimates. As a workaround for functional limitations of \texttt{pypngb}, we have estimated the lepton partner production cross sections as being entirely due to on-shell intermediate $W$ and $Z$ boson states, which then decay into one or more lepton partners:
\begin{equation}
    \sigma (pp \rightarrow f) \approx \sum_{i, X} \sigma (pp \rightarrow W_i) BR (W_i \rightarrow f+X) + 2 \sum_{j} \sigma (pp \rightarrow Z_j) BR (Z_j \rightarrow f\bar{f})
\end{equation}
for lepton resonances $f \in \{N_4, L_4, E2\}$. 

In general, the LM4DCHM$^{5-5-5}_{5-5}$ predicts lepton resonances that lie far beyond current and future collider sensitivity, necessitating alternative methods, such as indirect detection, for probing the models. The LM4DCHM$^{5-5-5}_{14-10}$ offers more accessible phenomenology, with $N_4$ featuring the highest predicted production cross-section among lepton signatures. In this model, the majority of points with masses ranging from $3$ to $6$~TeV exhibit predicted cross-sections in the range of 10$^{-12}$ to 10$^{-7}$~pb. By considering mass regions where these resonances would be produced at an integrated luminosity equivalent to the full HL-LHC run of $3000$~fb$^{-1}$, masses just below $3$~TeV will be produced by the end of HL-LHC, potentially enabling direct detection, especially for masses around $1$~TeV. However, it is important to note that we currently lack direct bounds from neutral lepton searches, as the available searches at $\sqrt{s}=13$~TeV either target mass ranges much lower than the predicted $N_4$ masses in both models or are limited to first and second-generation lepton partners.

Searches for heavy $\tau$ or vector-like leptons have been carried out in Refs.~\cite{ATLAS:2023vll,CMS:2019vll}, though their constraints are unable to rule out any of the existing points presented in our results. A small subset of the points in LM4DCHM$^{5-5-5}_{14-10}$ with masses around $1$ TeV may still be visible in direct particle searches, though this projection is optimistic as it does not account for background. Results from Ref.~\cite{Bhattiprolu_2019} suggest that only the lighter $L_4$ particles in the LM4DCHM$^{5-5-5}_{14-10}$ with $m_{L_4} < 1.25$~TeV will be excluded by the time HL-LHC has completed its full run, which should be sufficient to test the resonance band present in the LM4DCHM$^{5-5-5}_{14-10}$ at roughly $1.25$ TeV.

Searches for multiply-charged leptons have been conducted by ATLAS and CMS in Refs.~\cite{ATLAS:2023E2,CMS:2016kce}. Note that the ATLAS analysis places stricter bounds on $E2$ as their dataset is at a larger integrated luminosity and included photon-fusion processes, which are absent in both the CMS analysis and our calculation of the cross-sections. 

To summarise, most of the valid points highlighted by our global fits, including all those from the LM4DCHM$^{5-5-5}_{5-5}$, evade current collider sensitivity and furthermore would not even appear in direct particle searches by the end of the HL-LHC run. All of this is to say that the main avenue for testing the composite Higgs scenario might be to search for indirect evidence, such as through precision Higgs measurements, discussed below.

\subsection{Higgs signal strengths}
\label{exp_sig:higgs_sig_strengths}

\begin{figure}[h]
\centering
\begin{subfigure}{.49\linewidth}
  \centering
  \includegraphics[width=1\textwidth]{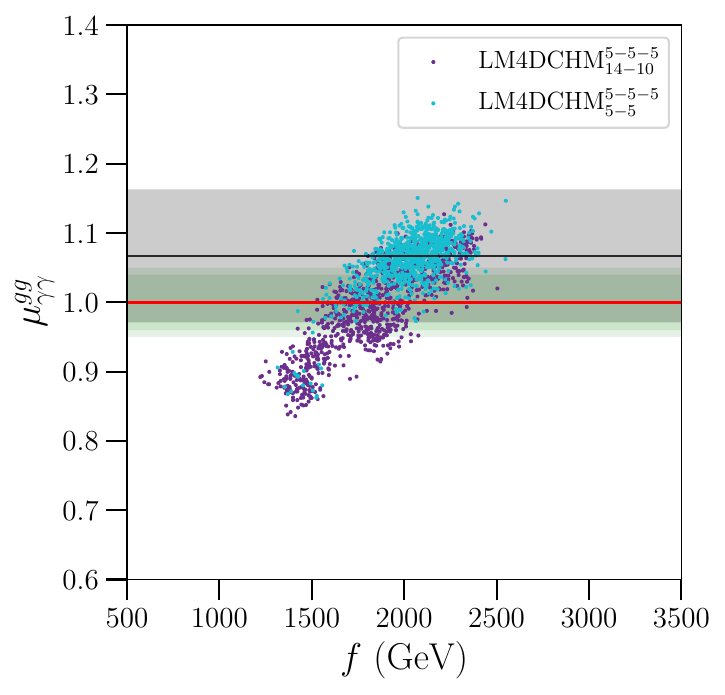}
\end{subfigure}
\begin{subfigure}{.49\linewidth}
  \centering
  \includegraphics[width=1\textwidth]{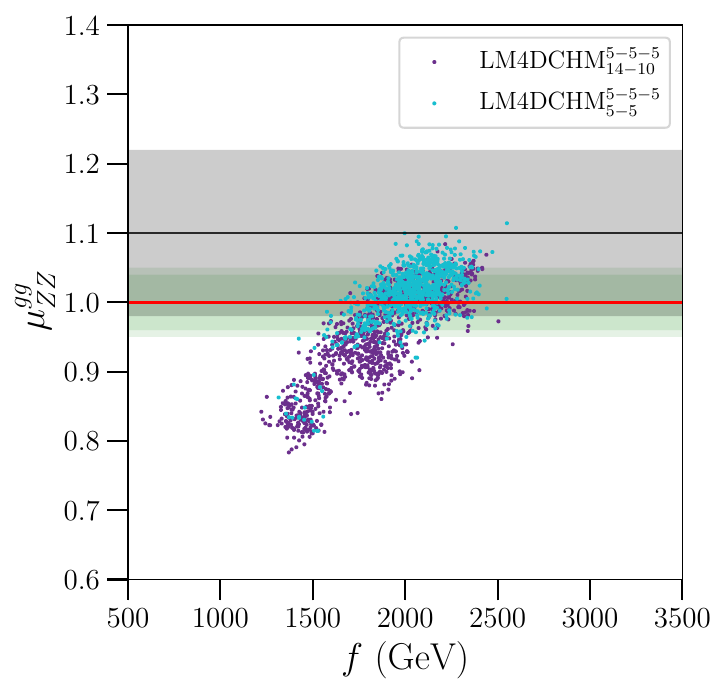}
\end{subfigure}  
\begin{subfigure}{.49\linewidth}
  \centering
  \includegraphics[width=1\textwidth]{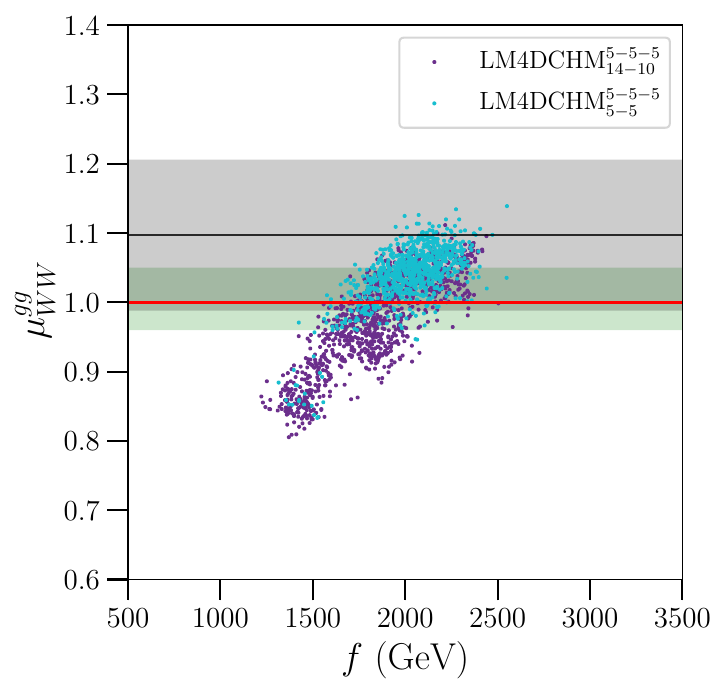}
\end{subfigure}
\begin{subfigure}{.49\linewidth}
  \centering
  \includegraphics[width=1\textwidth]{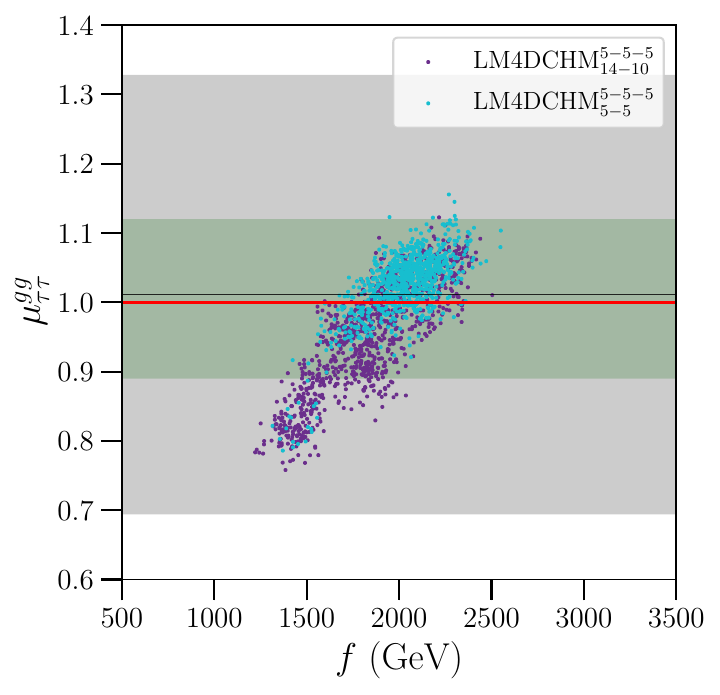}
\end{subfigure}
\caption{Higgs signal strengths for valid points in both the LM4DCHM$^{5-5-5}_{5-5}$ and LM4DCHM$^{5-5-5}_{14-5}$. Red lines show SM predictions for each decay channel, and the black lines and grey shaded areas show the experimentally measured values and their $1 \sigma$ uncertainties. Green shaded areas show the projected precision of measurements from the HL-LHC at 3000 fb$^{-1}$, assuming measurements centred on the SM value of $1$.}
\label{fig:H_sigstrengths}
\end{figure}

The predicted signal strengths from gluon-fusion produced Higgs decaying into pairs of $\gamma$, $W$ and $Z$ bosons are shown in \Cref{fig:H_sigstrengths}. Higgs signal strengths are excellent tests of CHMs, as these observables are sensitive to modifications of Higgs couplings to SM gauge bosons and fermions, as well as loop contributions from composite resonances. 

We see very similar predictions for each signal strength across the two models. This is to be expected as the Higgs is mainly coupled to both SM and composite gauge bosons, as well as $t$ and $b$ quarks --- hence varying the lepton representations is unlikely to alter the Higgs signal strengths. However, when comparing these results to those predicted by the quark-only M4DCHM$^{5-5-5}$ from \cite{Ethan}, we see that the inclusion of the third-generation leptons has removed the branches of signal strengths between $\sim 0.8-0.95$ that were roughly constant in $f$. Furthermore, predictions that align with experimental measurements for both models seem to occupy regions of $f$ slightly larger than those in the M4DCHM$^{5-5-5}$, now extending to values 2.25~TeV $\lesssim f \lesssim 2.5$~TeV.

Predictions of $\mu^{gg}_{\gamma \gamma}$ from both models broadly agree with both the SM value and experiment. Most valid points in the LM4DCHM$^{5-5-5}_{14-10}$ have signal strengths concentrated between $1$ and $1.1$, but still with a significant portion below this, while valid points for the LM4DCHM$^{5-5-5}_{5-5}$ center around ${\sim}1.09$. Furthermore, recent measurements of $\mu^{gg}_{\gamma \gamma}$ from CMS not included in our scans, $\mu^{gg}_{\gamma \gamma} = 1.07^{+0.12}_{-0.11}$ \cite{CMS:2021Hgaga}, show good agreement with our results (following from the already good agreement with the employed constraint).

The same is true for $\mu^{gg}_{WW}$ and $\mu^{gg}_{ZZ}$. Both models' predictions agree with experimental measurements, with the LM4DCHM$^{5-5-5}_{5-5}$ predicting signal strengths largely between $1$ and $1.1$, while the LM4DCHM$^{5-5-5}_{14-10}$ has values between $0.8$ and $1.1$. Similar to the diphoton signal strength, more recent measurements of the Z and W boson signal strengths from ATLAS and CMS \cite{CMS:2022uhn,CMS:2021ugl,CMS:2019chr,ATLAS:2022fnp,ATLAS:2019jst}, not directly included in our scans, have all yielded results compatible with the predictions of our models.

It is stressed again that valid points from the LM4DCHM$^{5-5-5}_{5-5}$ scan with the secondary mode at a larger $f$ are definitively more favoured by direct collider constraints, compared to other regions of the model. This smaller mode accounts entirely for the orange points in \Cref{fig:H_sigstrengths} that predict signal strengths near the experimentally measured values of around $1.1$. Indeed, the scan \textit{without} the smaller posterior mode only predicted points $1.1$~TeV~$< f \lesssim 1.5$~TeV with signal strengths $<1$ --- serving as a testament to how fine-tuned this particular model is.

As for the LM4DCHM$^{5-5-5}_{14-10}$, theoretically more natural values of $1$~TeV~$< f \lesssim 1.4$~TeV only predict signal strengths that are typically lower than experimentally observed values, indicating that this region is less favourable. The LM4DCHM$^{5-5-5}_{14-10}$ model's points are also concentrated at a slightly higher value of $f$ than the LM4DCHM$^{5-5-5}_{5-5}$ despite having the larger $f$ range of the two models.

All of this is likely due to the lack of double-tuning present in the LM4DCHM$^{5-5-5}_{14-10}$, whereas parameters in the LM4DCHM$^{5-5-5}_{5-5}$ need to take narrow values to reproduce electroweak symmetry breaking while also satisfying the imposed constraints. Any future work looking into different configurations of representations, particularly with quarks in the \textbf{14} and leptons included, would show which of these models might also be fine-tuned or outright excluded by experimental constraints. 

Prospects for Higgs signal strength measurements present promising opportunities to indirectly test both LM4DCHMs. Of particular significance are the signal strength measurements $\mu^{gg}_{\gamma \gamma}$, $\mu^{gg}_{WW}$, and $\mu^{gg}_{ZZ}$. ATLAS estimates that the uncertainties in these channels will be around $4$\%, while CMS predicts a $5$\% uncertainty\footnote{This prediction by CMS assumes the best-case scenario (S2+), where theoretical and systematic uncertainties are scaled down and accounting for all detector upgrades; the ATLAS prediction assumes no theoretical uncertainty.} \cite{ATLAS:2018jlh,CMS-PAS-FTR-16-002}. The precision of $\mu^{gg}_{\tau \tau}$ is also expected to improve substantially, although not enough to distinguish between the two models through this channel. Similarly, the precision of $\mu^{gg}_{b b}$ measurements is expected to improve, prompting us to consider including this channel in future scans. 

\Cref{fig:H_sigstrengths} illustrates these projections centered on the SM value of $1$, and demonstrates that such precision will serve as an relatively strong test of these models, in particular the viability of the LM4DCHM$^{5-5-5}_{5-5}$.

%% file: tex/Appendix.tex
\section{Lepton mass matrices}
\label{appendix_mass_matrices}

Mass matrices for the bosons and quarks in our models can be found in the Appendices of Ref.~\cite{Ethan}. Since the lepton mass terms have the same form as those of the quarks, they will have the same mass matrices with the appropriate substitution of coefficients. For the benefit of readers, the expressions for the lepton mass matrices are provided below. Masses of the physical fields will be the singular values of these matrices.

The composite multiplets $\Psi^{\tau}$ and $\tilde{\Psi}^{\tau}$ contain fields $\Psi^{n_{L},n_{R}}$ and $\tilde{\Psi}^{n_{L},n_{R}}$ with $SU(2)_L \times SU(2)_R$ quantum numbers $(n_{L},n_{R})$, and therefore electric charges $n_{L} + n_{R} + Y$. The basis of fields in which the mass matrices are written will be indicated alongside each matrix. Since the $SO(5)$ representations branch into $SO(4)$ representations as
\begin{align}
    \mathbf{5} &\rightarrow \mathbf{4} \oplus \mathbf{1}, \nonumber\\
    \mathbf{10} &\rightarrow \mathbf{6} \oplus \mathbf{4}, \nonumber\\
    \mathbf{14} &\rightarrow \mathbf{9} \oplus \mathbf{4} \oplus \mathbf{1},
\end{align}
the component fields $\Psi^{n_{L},n_{R}}$ will be given a subscript to indicate which $SO(4)$ representation they belong to where necessary. We will write $n_{L,R} = \pm$ to indicate $n_{L,R} = \pm 1/2$ for fields in the $\mathbf{4}$ representation, while for the other representations this will indicate $n_{L,R} = \pm 1$. 

\noindent The matrices will make use of the quantities
\begin{align}
    s_{xh} = \sin \brackets{x\frac{h}{f}},\quad c_{xh} = \cos \brackets{x\frac{h}{f}} \quad \text{for } x \in \mathbb{R},
\end{align}
and
\begin{align}
    c_{\pm} = \frac{c_{h} \pm c_{2h}}{2},\ \tilde{c} = \frac{3 + 5 c_{2h}}{8},  \text{ and } \tilde{s} = \frac{\sqrt{5}}{4} s_{2h}.\\
\end{align}

\subsection{LM4DCHM$_{5-5}$}

In this model the composite charge -2 resonances have mass matrix
\begin{align}
M_{M2} = \left(\begin{array}{c|cc}
    {} & \Psi^{-,-}_{R} & \tilde{\Psi}^{-,-}_{R}  \\[2pt]
    \hline\\\\[-4.5\medskipamount]
    \bar{\Psi}^{-,-}_{L} & m_\tau & m_{Y_{\tau}}  \\
    \bar{\tilde{\Psi}}^{-,-}_{L} &  0 &  m_{\tilde{\tau}} \end{array}\right),
\end{align}
while the neutral resonances have mass matrix
\begin{align}
M_{N}  = \left(\begin{array}{c|ccc}
    {} & \nu^{0}_{R} & \Psi^{+,+}_{R} & \tilde{\Psi}^{+,+}_{R}  \\[2pt]
    \hline\\\\[-4.5\medskipamount]
    \bar{\nu}^{0}_{L} & 0 & -\Delta_{\tau L} & 0 \\
    \bar{\Psi}^{+,+}_{L} & 0 & m_\tau & m_{Y_{\tau}}  \\
    \bar{\tilde{\Psi}}^{+,+}_{L} & 0 & 0 &  m_{\tilde{\tau}} \end{array}\right).
\end{align}
Notice that since the right-handed neutrino is not included in the model, its couplings are all zero and we are guaranteed a massless neutral state - the SM neutrino. Finally, the charge -1 mass matrix is
\begin{align}
M_{L} = \left( \begin{array}{c|ccccccc}
    {} & \tau^0_R & \Psi^{+,-}_{R} & \tilde{\Psi}^{+,-}_{R} & \Psi^{-,+}_{R} & \tilde{\Psi}^{-,+}_{R}  & \Psi^{0,0}_{R} & \tilde{\Psi}^{0,0}_{R} \\[2pt]
    \hline\\\\[-4.5\medskipamount]
    \bar{\tau}^0_L & 0 & s^{2}_{h/2} \Delta_{\tau L} & 0 & - c^{2}_{h/2} \Delta_{\tau L} & 0 & \frac{i}{\sqrt{2}} s_{h} \Delta_{\tau L} & 0 \\
    \bar{\Psi}^{+,-}_{L} & 0 & m_{\tilde{\tau}} & m_{Y_\tau} & 0 & 0 & 0 & 0 \\
    \bar{\tilde{\Psi}}^{+,-}_{L} & - \frac{i}{\sqrt{2}} s_{h} \Delta^{\dagger}_{\tau R} & 0 & m_{\tilde{\tau}} & 0 & 0 & 0 & 0 \\
    \bar{\Psi}^{-,+}_{L} & 0 & 0 & 0 & m_{\tilde{\tau}} & m_{Y_\tau} & 0 & 0 \\
    \bar{\tilde{\Psi}}^{-,+}_{L} &  - \frac{i}{\sqrt{2}} s_{h} \Delta^{\dagger}_{\tau R} & 0 & 0 & 0 & m_{\tilde{\tau}} & 0 & 0 \\
    \bar{\Psi}^{0,0}_{L} & 0 & 0 & 0 & 0 & 0 & m_{\tilde{\tau}} & m_{Y_\tau} + Y_{\tau} \\
    \bar{\tilde{\Psi}}^{0,0}_{L} & - c_{h} \Delta^{\dagger}_{\tau R} & 0 & 0 & 0 & 0 & 0 & m_{\tilde{\tau}}
    \end{array}\right).
\end{align}

\subsection{LM4DCHM$_{14-10}$}

This model contains a single charge +2 field and a single charge -2 field, both of which have mass $m_{\tau}$. There are also charge +1 fields, which have a mass matrix
\begin{align}
M_{E1} = \left(\begin{array}{c|cccccc}
    {} & \tilde{\Psi}^{+,+}_R & \Psi^{+,+}_R & \Psi^{+,0}_R & \Psi^{0,+}_R & \tilde{\Psi}^{+,0}_R & \tilde{\Psi}^{0,+}_R  \\[2pt]
    \hline\\\\[-4.5\medskipamount]
    \bar{\tilde{\Psi}}^{+,+}_L & m_{\tilde{\tau}} & 0 & 0 & 0 & 0 & 0 \\[1pt]
    \bar{\Psi}^{+,+}_L & \frac{1}{2} Y_{\tau} & m_{\tau} & 0 & 0 & 0 & 0 \\[1pt]
    \bar{\Psi}^{+,0}_L & 0 & 0 & m_{\tau} & 0 & 0 & 0 \\[1pt]
    \bar{\Psi}^{0,+}_L & 0 & 0 & 0 & m_{\tau} & 0 & 0 \\[1pt]
    \bar{\tilde{\Psi}}^{+,0}_L & 0 & 0 & 0 & 0 & m_{\tilde{\tau}} & 0 \\[1pt]
    \bar{\tilde{\Psi}}^{0,+}_L & 0 & 0 & 0 & 0 & 0 & m_{\tilde{\tau}} \end{array}\right).
\end{align}
The charge -1 mass matrix is
\begin{align}
&M_{L} = \left(\begin{array}{c|ccccccc}
    {} & \tau^0_R & \tilde{\Psi}^{-,-}_R & \Psi^{-,-}_R & \Psi^{-,0}_R & \Psi^{0,-}_R & \tilde{\Psi}^{-,0}_R & \tilde{\Psi}^{0,-}_R \\[2pt]
    \hline\\\\[-4.5\medskipamount]
     \bar{\tau}^0_L & 0 & 0 & -c_{h} \Delta_{\tau L} & -\frac{1}{\sqrt{2}} s_{h} \Delta_{\tau L} & \frac{i}{\sqrt{2}} s_{h} \Delta_{\tau L} & 0 & 0 \\[1pt]
     \bar{\tilde{\Psi}}^{-,-}_L & -\frac{i}{\sqrt{2}} s_{h} \Delta^{\dagger}_{\tau R} & m_{\tilde{\tau}} & 0 & 0 & 0 & 0 & 0 \\[1pt]
     \bar{\Psi}^{-,-}_L & 0 & \frac{1}{2} Y_{\tau} & m_{\tau} & 0 & 0 & 0 & 0 \\[1pt]
     \bar{\Psi}^{-,0}_L & 0 & 0 & 0 & m_{\tau} & 0 & 0 & 0 \\[1pt]
     \bar{\Psi}^{0,-}_L & 0 & 0 & 0 & 0 & m_{\tau} & 0 & 0 \\[1pt]
     \bar{\tilde{\Psi}}^{-,0}_L & - s^{2}_{h/2} \Delta^{\dagger}_{\tau R} & 0 & 0 & 0 & 0 & m_{\tilde{\tau}} & 0 \\[1pt]
     \bar{\tilde{\Psi}}^{0,-}_L & - c^{2}_{h/2} \Delta^{\dagger}_{\tau R} & 0 & 0 & 0 & 0 & 0 & m_{\tilde{\tau}} \end{array}\right),\\
\end{align}
and the neutral resonance mass matrix is
\begin{align}
&M_{N} =\nonumber\\
&\quad \left(\begin{array}{c|ccccccccccc}
    {} & \nu^0_R & \Psi^{0,0}_{\mathbf{1}_R} & \tilde{\Psi}^{+,-}_R & \Psi^{+,-}_{\mathbf{4}_R} & \tilde{\Psi}^{-,+}_R & \Psi^{-,+}_{\mathbf{4}_R} & \Psi^{-,+}_{\mathbf{9}_R} & \Psi^{0,0}_{\mathbf{9}_R} & \Psi^{+,-}_{\mathbf{9}_R} & \tilde{\Psi}^{0,0}_{1_R} & \tilde{\Psi}^{0,0}_{2_R} \\[2pt]
    \hline\\\\[-4.5\medskipamount]
     \bar{\nu}^0_L & 0 & \tilde{s} \Delta_{\tau L} & 0 & - c_{+} \Delta_{\tau L} & 0 & -i c_{-} \Delta_{\tau L} & s_{h} s^{2}_{h/2} \Delta_{\tau L} & -\frac{s_{2h}}{4} \Delta_{\tau L} & s_{h} c^{2}_{h/2} \Delta_{\tau L} & 0 & 0 \\[1pt]
     \bar{\Psi}^{0,0}_{\mathbf{1}_L} & 0 & m_{\tau} & 0 & 0 & 0 & 0 & 0 & 0 & 0 & 0 & 0 \\[1pt]
     \bar{\tilde{\Psi}}^{+,-}_L & 0 & 0 & m_{\tilde{\tau}} & 0 & 0 & 0 & 0 & 0 & 0 & 0 & 0 \\[1pt]
     \bar{\Psi}^{+,-}_{\mathbf{4}_L} & 0 & 0 & \frac{1}{2} Y_{\tau} & m_{\tau} & 0 & 0 & 0 & 0 & 0 & 0 & 0 \\[1pt]
     \bar{\tilde{\Psi}}^{-,+}_L & 0 & 0 & 0 & 0 & m_{\tilde{\tau}} & 0 & 0 & 0 & 0 & 0 & 0 \\[1pt]
     \bar{\Psi}^{-,+}_{\mathbf{4}_L} & 0 & 0 & 0 & 0 & \frac{1}{2} Y_{\tau} & m_{\tau} & 0 & 0 & 0 & 0 & 0 \\[1pt]
     \bar{\Psi}^{-,+}_{\mathbf{9}_L} & 0 & 0 & 0 & 0 & 0 & 0 & m_{\tau} & 0 & 0 & 0 & 0 \\[1pt]
     \bar{\Psi}^{0,0}_{\mathbf{9}_L} & 0 & 0 & 0 & 0 & 0 & 0 & 0 & m_{\tau} & 0 & 0 & 0 \\[1pt]
     \bar{\Psi}^{+,-}_{\mathbf{9}_L} & 0 & 0 & 0 & 0 & 0 & 0 & 0 & 0 & m_{\tau} & 0 & 0 \\[1pt]
     \bar{\tilde{\Psi}}^{0,0}_{1_L} & 0 & 0 & 0 & 0 & 0 & 0 & 0 & 0 & 0 & m_{\tilde{\tau}} & 0 \\[1pt]
     \bar{\tilde{\Psi}}^{0,0}_{2_L} & 0 & 0 & 0 & 0 & 0 & 0 & 0 & 0 & 0 & 0 & m_{\tilde{\tau}} \end{array}\right).
\end{align}

\section{Form factors}
\label{appendix_correlators}

This section provides expressions for the form factors $\Pi$ and $M$ of \Cref{eq:Higgs_potentialFF}. In both models, the form factors are given in terms of the functions
\begin{align}
    & A_{R}(m_{1},m_{2},m_{3},m_{4},\Delta) := \Delta^{2} (m_{1}^{2} m_{2}^{2} + m_{2}^{2} m_{3}^{2} - p^{2} (m_{1}^{2} + m_{2}^{2} + m_{3}^{2} + m_{4}^{2}) + p^{4}), \nonumber\\
    & A_{L}(m_{1},m_{2},m_{3},m_{4},\Delta) := \Delta^{2} m_{1}^{2} m_{4}^{2} + A_{R}(m_{1},m_{2},m_{3},m_{4},\Delta), \nonumber\\
    & A_{M}(m_{1},m_{2},m_{3},m_{4},\Delta_{1},\Delta_{2}) := \Delta_{1} \Delta_{2} m_{1} m_{2} m_{4} (m_{3}^{2} - p^{2}), \nonumber\\
    & B(m_{1},m_{2},m_{3},m_{4},m_{5}) := m_{1}^{2} m_{2}^{2} m_{3}^{2} - p^{2} (m_{1}^{2} m_{2}^{2} + m_{1}^{2} m_{3}^{2} + m_{2}^{2} m_{3}^{2} + m_{2}^{2} m_{5}^{2} + m_{3}^{2} m_{4}^{2}) \nonumber\\
    &\hphantom{B(m_{1},m_{2},m_{3},m_{4},m_{5}) :=} + p^{4} (m_{1}^{2} + m_{2}^{2} + m_{3}^{2} + m_{4}^{2} + m_{5}^{2}) - p^{6},
\end{align}
in Minkowski space.

\subsection*{Quark sector}

Both models share a $\mathbf{5}-\mathbf{5}-\mathbf{5}$ quark sector, for which the form factors are
\begin{align}
    \Pi_{t_{L}} &= \Pi^{(4)}_{q_{t}} + \Pi^{(4)}_{q_{b}} + \frac{1}{2} \brackets{\Pi^{(1)}_{q_{t}} - \Pi^{(4)}_{q_{t}}} s^{2}_{h}, && M_{t} = \brackets{M^{(1)}_{t} - M^{(4)}_{t}} \sqrt{\frac{1-s^{2}_{h}}{2}} s_{h}, \nonumber\\
    \Pi_{t_{R}} &= \Pi^{(1)}_{t} - \brackets{\Pi^{(1)}_{t} - \Pi^{(4)}_{t}} s^{2}_{h},
\end{align}
where
\begin{align}
    \Pi^{(1)}_{q_{t}} &= \frac{A_{L}(m_{\tilde{t}},0,m_{Y_{t}} + Y_{t},0,\Delta_{tL})}{B(m_{t},m_{\tilde{t}},0,m_{Y_{t}} + Y_{t},0)},  && \Pi^{(4)}_{q_{t}} = \frac{A_{L}(m_{\tilde{t}},0,m_{Y_{t}},0,\Delta_{tL})}{B(m_{t},m_{\tilde{t}},0,m_{Y_{t}},0)}, \nonumber\\
    \Pi^{(1)}_{t} &= \frac{A_{L}(m_{t},0,m_{Y_{t}} + Y_{t},0,\Delta_{tR})}{B(m_{t},m_{\tilde{t}},0,m_{Y_{t}} + Y_{t},0)},  && \Pi^{(4)}_{t} = \frac{A_{L}(m_{t},0,m_{Y_{t}},0,\Delta_{tR})}{B(m_{t},m_{\tilde{t}},0,m_{Y_{t}},0)}, \nonumber\\
    M^{(1)}_{t} &= \frac{A_{M}(m_{t},m_{\tilde{t}},0,m_{Y_{t}} + Y_{t},\Delta_{tL},\Delta_{tR})}{B(m_{t},m_{\tilde{t}},0,m_{Y_{t}} + Y_{t},0)},  && M^{(4)}_{t} = \frac{A_{M}(m_{t},m_{\tilde{t}},0,m_{Y_{t}},\Delta_{tL},\Delta_{tR})}{B(m_{t},m_{\tilde{t}},0,m_{Y_{t}},0)}. \nonumber\\
\end{align}
Interchanging all $t$ and $\tilde{t}$ subscripts with $b$ and $\tilde{b}$ yields the form factors for the bottom sector.

\subsection{LM4DCHM$^{5-5-5}_{5-5}$}

In this model, the lepton sector is
\begin{align}
    \Pi_{\tau_{L}} &= \Pi^{(4)}_{l_{\tau}} + \Pi^{(4)}_{l_{b}} + \frac{1}{2} \brackets{\Pi^{(1)}_{l_{\tau}} - \Pi^{(4)}_{l_{\tau}}} s^{2}_{h}, && M_{\tau} = \brackets{M^{(1)}_{\tau} - M^{(4)}_{\tau}} \sqrt{\frac{1-s^{2}_{h}}{2}} s_{h}, \nonumber\\
    \Pi_{\tau_{R}} &= \Pi^{(1)}_{\tau} - \brackets{\Pi^{(1)}_{\tau} - \Pi^{(4)}_{\tau}} s^{2}_{h},
\end{align}
where
\begin{align}
    \Pi^{(1)}_{l_{\tau}} &= \frac{A_{L}(m_{\tilde{\tau}},0,m_{Y_{\tau}} + Y_{\tau},0,\Delta_{\tau L})}{B(m_{\tau},m_{\tilde{\tau}},0,m_{Y_{\tau}} + Y_{\tau},0)},  && \Pi^{(4)}_{l_{\tau}} = \frac{A_{L}(m_{\tilde{\tau}},0,m_{Y_{\tau}},0,\Delta_{\tau L})}{B(m_{\tau},m_{\tilde{\tau}},0,m_{Y_{\tau}},0)}, \nonumber\\
    \Pi^{(1)}_{\tau} &= \frac{A_{L}(m_{\tau},0,m_{Y_{\tau}} + Y_{\tau},0,\Delta_{\tau R})}{B(m_{\tau},m_{\tilde{\tau}},0,m_{Y_{\tau}} + Y_{\tau},0)},  && \Pi^{(4)}_{\tau} = \frac{A_{L}(m_{\tau},0,m_{Y_{\tau}},0,\Delta_{\tau R})}{B(m_{\tau},m_{\tilde{\tau}},0,m_{Y_{\tau}},0)}, \nonumber\\
    M^{(1)}_{\tau} &= \frac{A_{M}(m_{\tau},m_{\tilde{\tau}},0,m_{Y_{\tau}} + Y_{\tau},\Delta_{\tau L},\Delta_{\tau R})}{B(m_{\tau},m_{\tilde{\tau}},0,m_{Y_{\tau}} + Y_{\tau},0)},  && M^{(4)}_{\tau} = \frac{A_{M}(m_{\tau},m_{\tilde{\tau}},0,m_{Y_{\tau}},\Delta_{\tau L},\Delta_{\tau R})}{B(m_{\tau},m_{\tilde{\tau}},0,m_{Y_{\tau}},0)}. \nonumber\\
\end{align}

\subsection{LM4DCHM$^{5-5-5}_{14-10}$}

In this model, the lepton sector is
\begin{align}
    \Pi_{\tau_{L}} &= \Pi^{(4)}_{l} - \brackets{\Pi^{(4)}_{l} - \Pi^{(9)}_{l}} s^{2}_{h}, && M_{\tau} = - i M^{(4)}_{\tau} s_{h} \sqrt{\frac{1-s^{2}_{h}}{2}}, \nonumber\\
    \Pi_{\tau_{R}} &= \Pi^{(6)}_{\tau} + \brackets{\Pi^{(4)}_{\tau} - \Pi^{(6)}_{\tau}} \frac{s^{2}_{h}}{2},
\end{align}
where
\begin{align}
    \Pi^{(10)}_{l} &= \frac{A_{L}(0,0,0,0,\Delta_{l})}{B(m_{l},0,0,0,0)}, && \Pi^{(6)}_{\tau} = \frac{A_{R}(0,0,0,0,\Delta_{\tau})}{B(0,m_{\tau},0,0,0)}, \nonumber\\
    \Pi^{(4)}_{l} &= \frac{A_{L}(0,m_{\tau},0,Y_{\tau}/2,\Delta_{l})}{B(m_{l},0,m_{\tau},0,Y_{\tau}/2)}, && \Pi^{(4)}_{\tau} = \frac{A_{R}(m_{l},0,Y_{\tau}/2,0,\Delta_{\tau})}{B(m_{l},0,m_{\tau},0,Y_{\tau}/2)}, \nonumber\\
     M^{(4)}_{\tau} &= - i \frac{A_{M}(m_{l},m_{\tau},0,Y_{\tau}/2,\Delta_{l},\Delta_{\tau})}{B(m_{l},0,m_{\tau},0,Y_{\tau}/2)}.
\end{align}

\section{Scan comparisons}
\label{scan_agreement_appendix}

\subsection{LM4DCHM$^{5-5-5}_{5-5}$}
\label{appendix:5-5}
\begin{figure}[h]
\centering
  \includegraphics[width=1\linewidth]{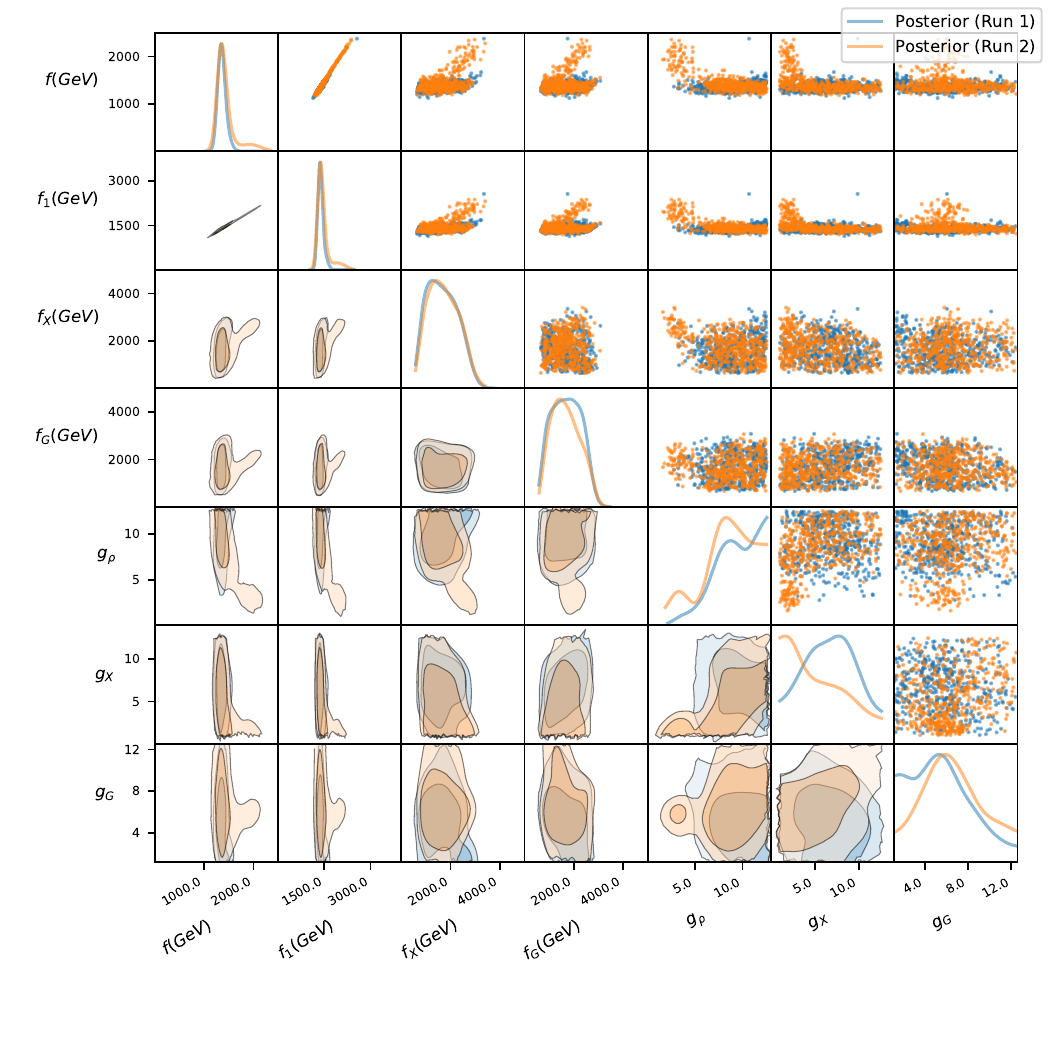}
\caption{1D and 2D marginalised posteriors for the gauge sector parameters in the LM4DCHM$^{5-5-5}_{5-5}$ found in two different runs with $4000$ live points.}
\label{fig:5-5_gauge_posteriors}
\end{figure}

\begin{figure}[h]
\centering
  \includegraphics[width=1\linewidth]{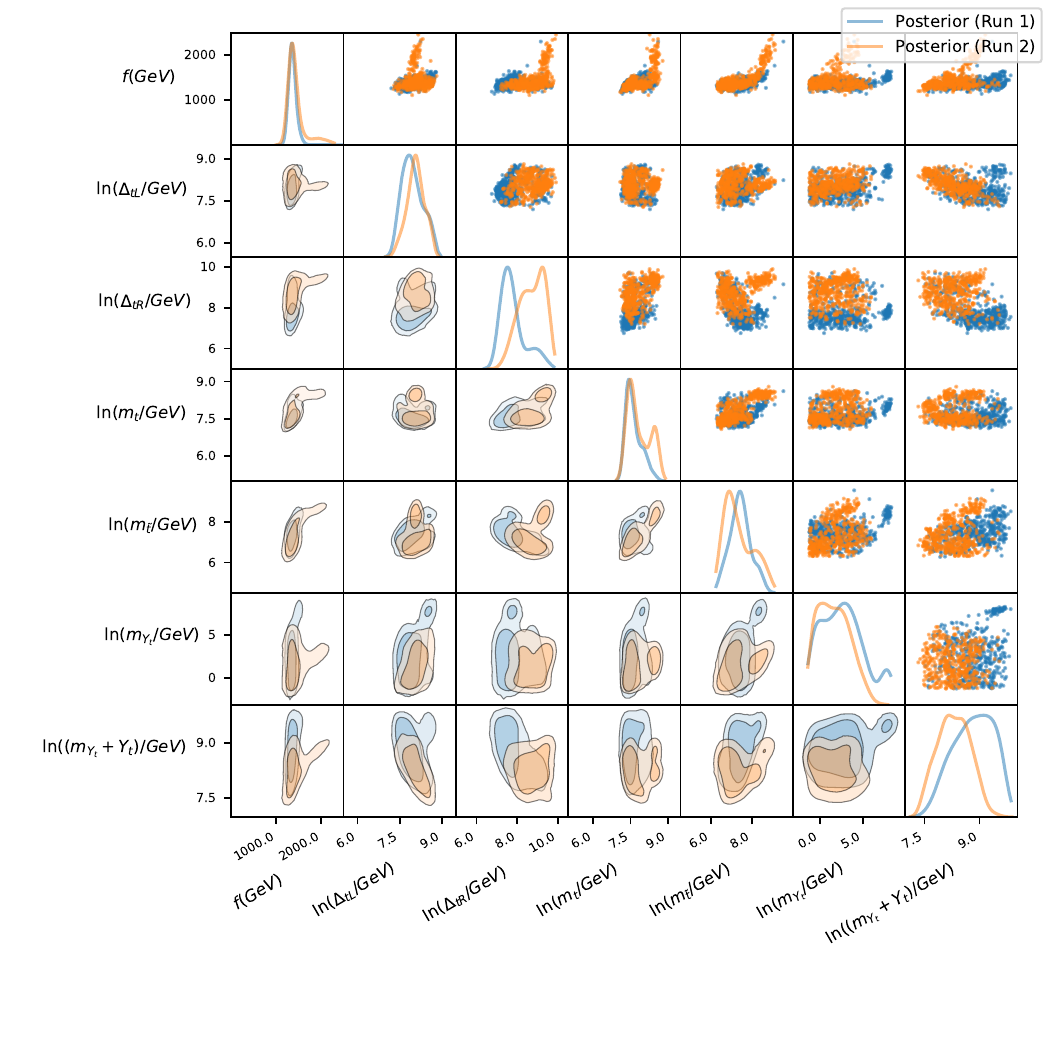}
\caption{1D and 2D marginalised posteriors for the top quark parameters in the LM4DCHM$^{5-5-5}_{5-5}$ found in two different runs with $4000$ live points. The decay constant $f$ has been included to highlight the extra posterior mode found in Run 2.}
\label{fig:5-5_f_v_top_posteriors}
\end{figure}

\begin{figure}[h]
\centering
  \includegraphics[width=1\linewidth]{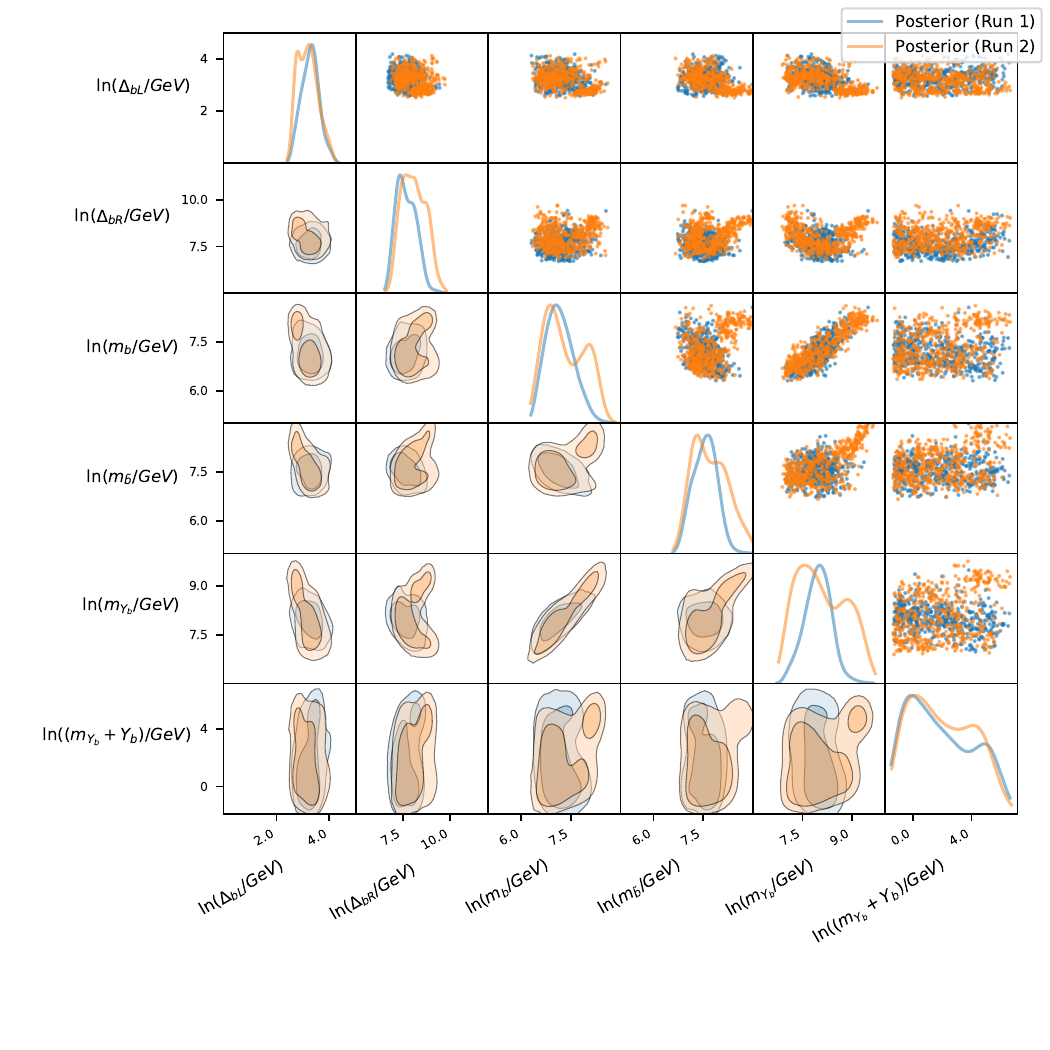}
\caption{1D and 2D marginalised posteriors for the bottom quark parameters in the LM4DCHM$^{5-5-5}_{5-5}$ found in two different runs with $4000$ live points.}
\label{fig:5-5_bottom_posteriors}
\end{figure}

\begin{figure}[h]
\centering
  \includegraphics[width=1\linewidth]{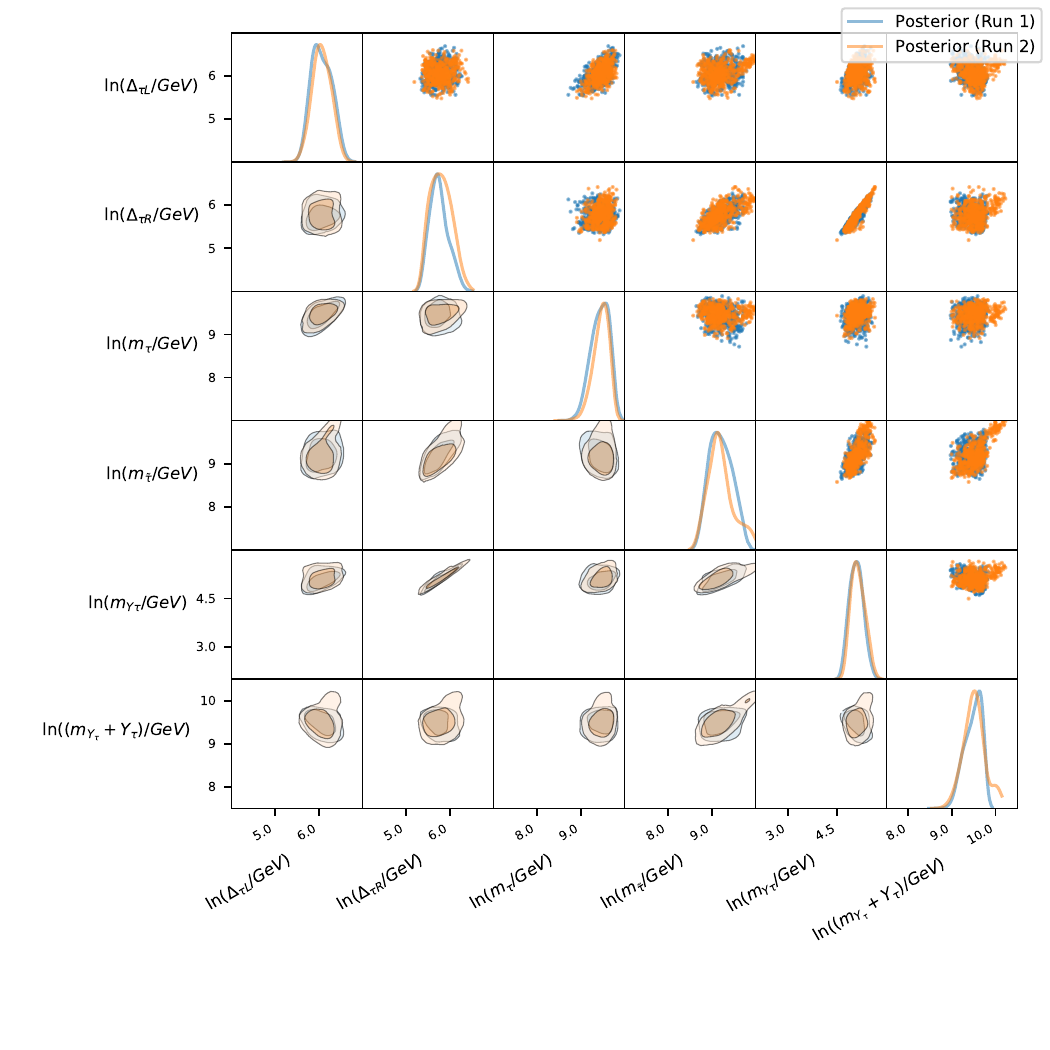}
\caption{1D and 2D marginalised posteriors for the tau lepton parameters in the LM4DCHM$^{5-5-5}_{5-5}$ found in two different runs with $4000$ live points.}
\label{fig:5-5_tau_posteriors}
\end{figure}

We performed two scans of the LM4DCHM$^{5-5-5}_{5-5}$ with 4000 live points each. The posterior distributions found in each scan are shown in \Cref{fig:5-5_gauge_posteriors,fig:5-5_f_v_top_posteriors,fig:5-5_bottom_posteriors,fig:5-5_tau_posteriors}. There tends to be good agreement in the posteriors between the two scans, except for some larger discrepancies in the top quark sector, which has always proven difficult for top partners in the $\mathbf{5}$ representation. There is excellent agreement in the lepton parameters, which are the focus of the present paper. The evidences found in the scans,
\begin{align}
    \ln(\mathcal{Z})_{\text{Run 1}} &= -45.54 \pm 0.08, \nonumber\\
    \ln(\mathcal{Z})_{\text{Run 2}} &= -44.85 \pm 0.08,
\end{align}
are in some tension with each other. However, they are of the same approximate size, and are much smaller than the evidences found for the LM4DCHM$^{5-5-5}_{14-10}$, so any small discrepancies in these values should not cast doubt on the conclusion that the LM4DCHM$^{5-5-5}_{14-10}$ is the greatly preferred model.

Only Run 2 found the extra posterior modes that contain most of the valid points that survive direct detector constraints.
Other than the secondary mode at $f \approx 2$~TeV, we can see from \Cref{fig:5-5_f_v_top_posteriors} that an additional mode is present where $3$~TeV~$\lesssim \Delta_{tR} \lesssim 22 $~TeV and $1.8$~TeV $\lesssim m_{Yt} + Y_t \lesssim 8$~TeV. These modes are responsible for the valid points mentioned in \Cref{exp_sig_section} that survive all constraints and collider bounds at $3 \sigma$.

\clearpage
\subsection{LM4DCHM$^{5-5-5}_{14-10}$}

\begin{figure}[h]
\centering
  \includegraphics[width=1\linewidth]{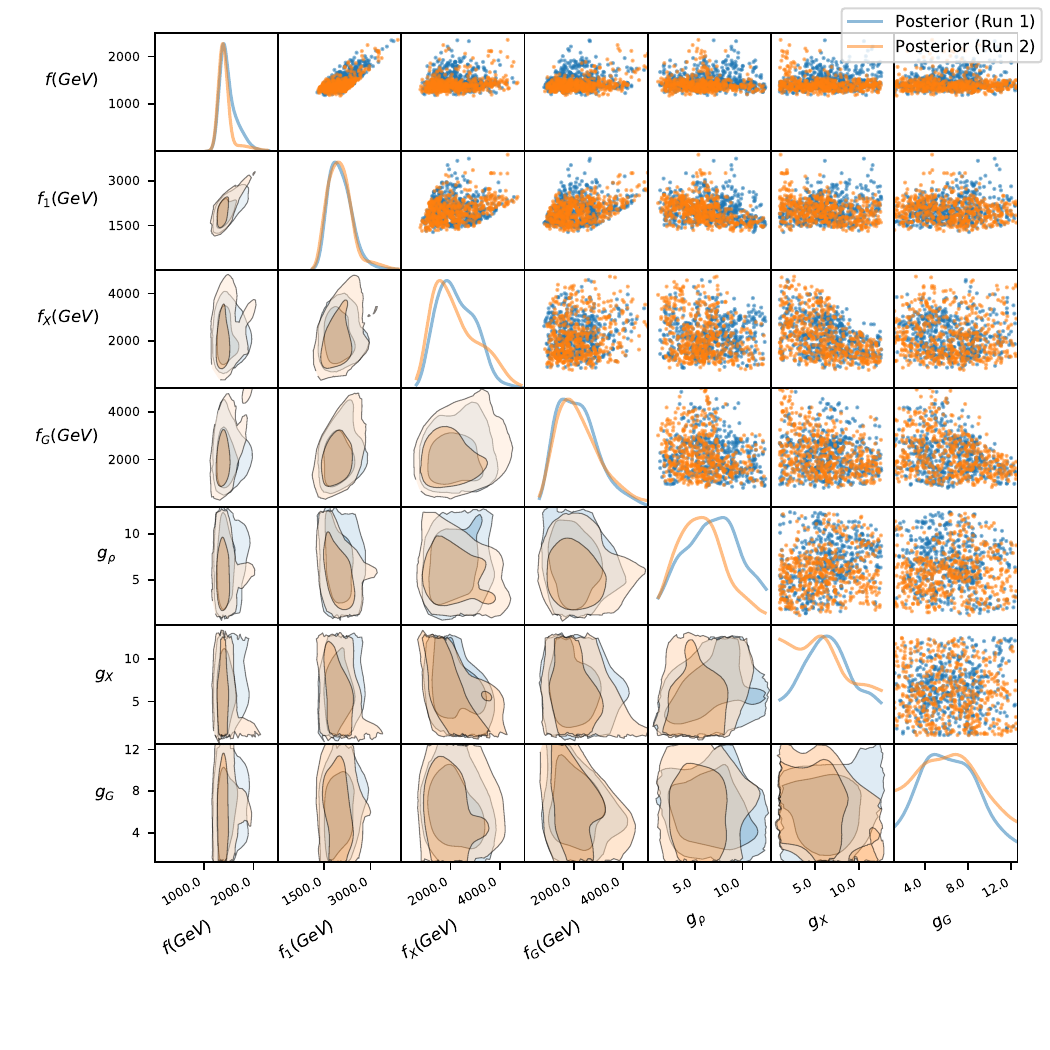}
\caption{1D and 2D marginalised posteriors for the gauge sector parameters in the LM4DCHM$^{5-5-5}_{14-10}$ found in two different runs with $4000$ live points.}
\label{fig:14-10_gauge_posteriors}
\end{figure}

\begin{figure}[h]
\centering
  \includegraphics[width=1\linewidth]{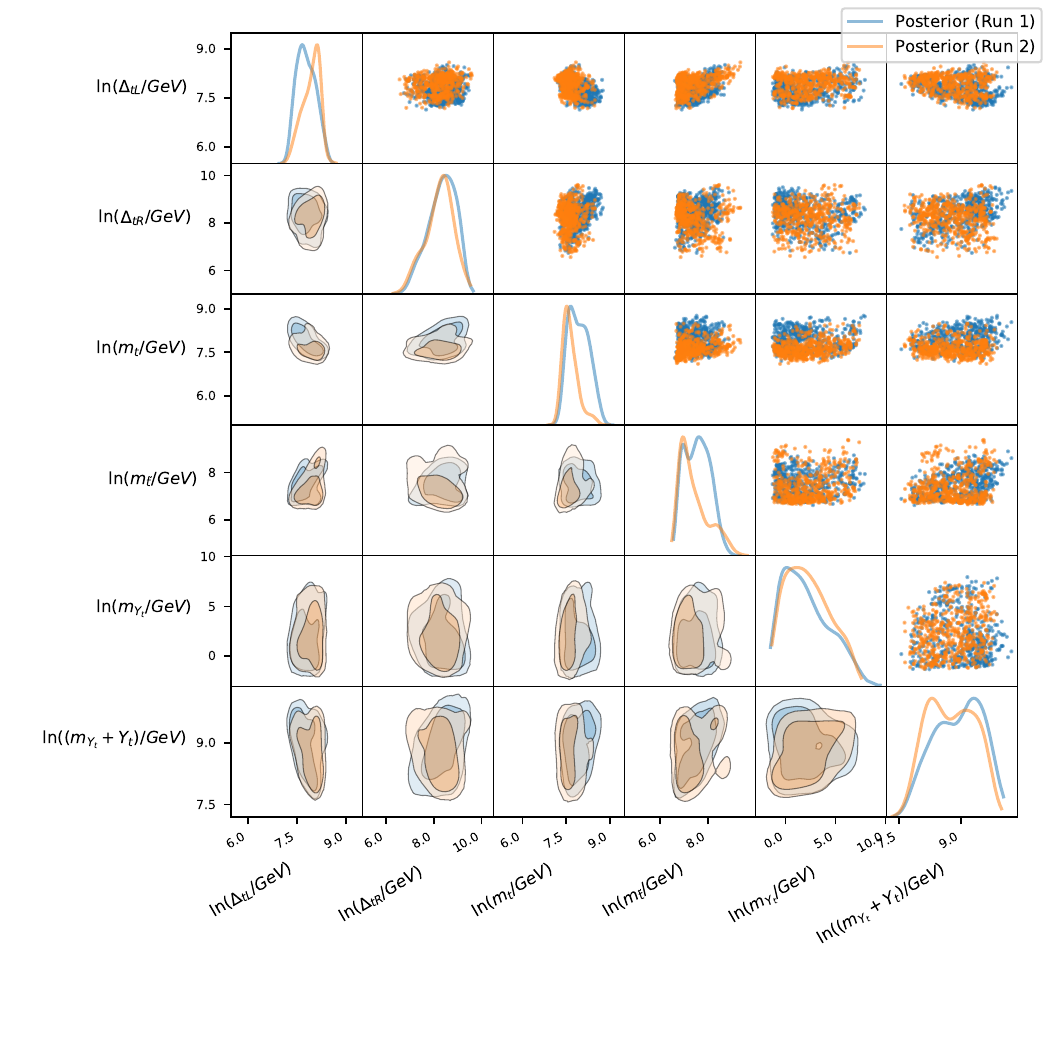}
\caption{1D and 2D marginalised posteriors for the top quark sector parameters in the LM4DCHM$^{5-5-5}_{14-10}$ found in two different runs with $4000$ live points.}
\label{fig:14-10_top_posteriors}
\end{figure}

\begin{figure}[h]
\centering
  \includegraphics[width=1\linewidth]{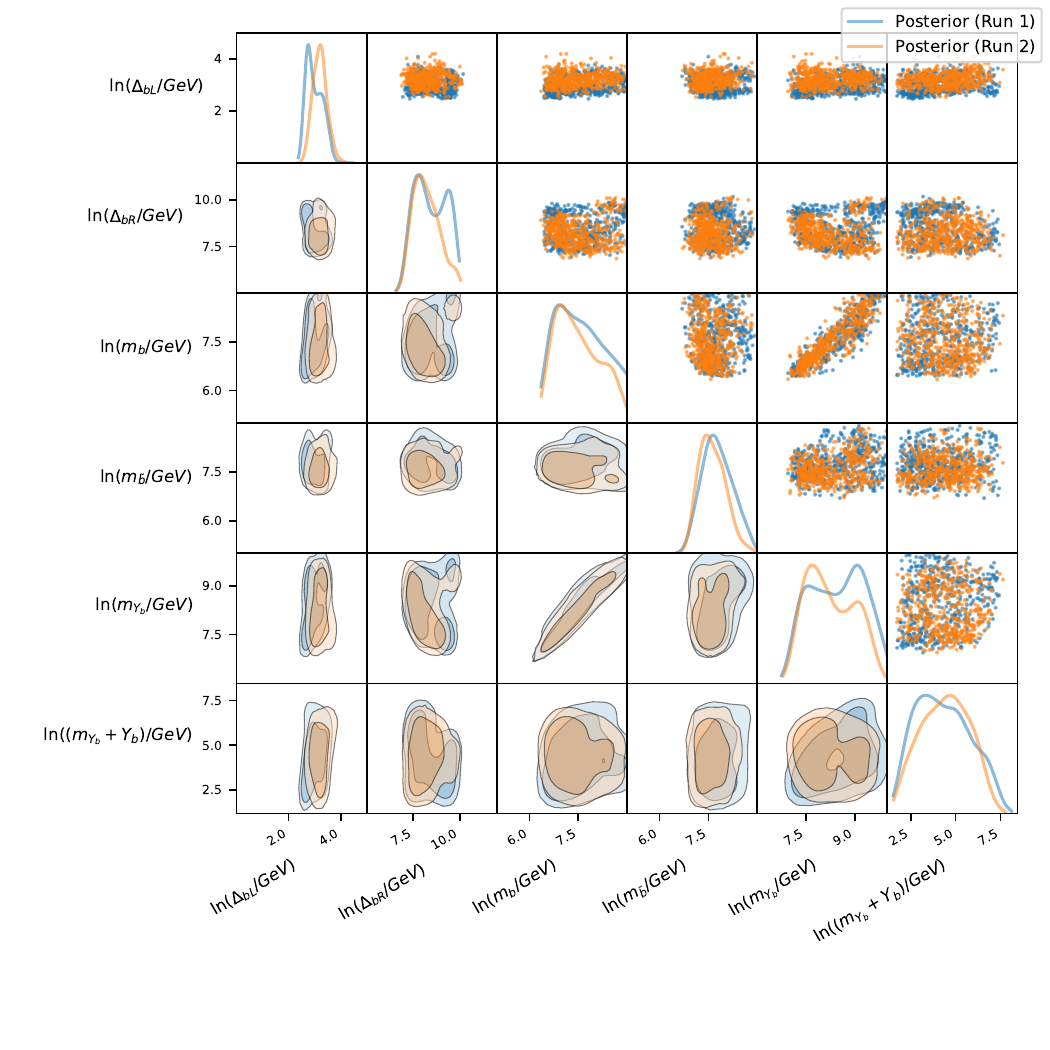}
\caption{1D and 2D marginalised posteriors for the bottom quark sector parameters in the LM4DCHM$^{5-5-5}_{14-10}$ found in two different runs with $4000$ live points.}
\label{fig:14-10_bottom_posteriors}
\end{figure}

\begin{figure}[h]
\centering
  \includegraphics[width=1\linewidth]{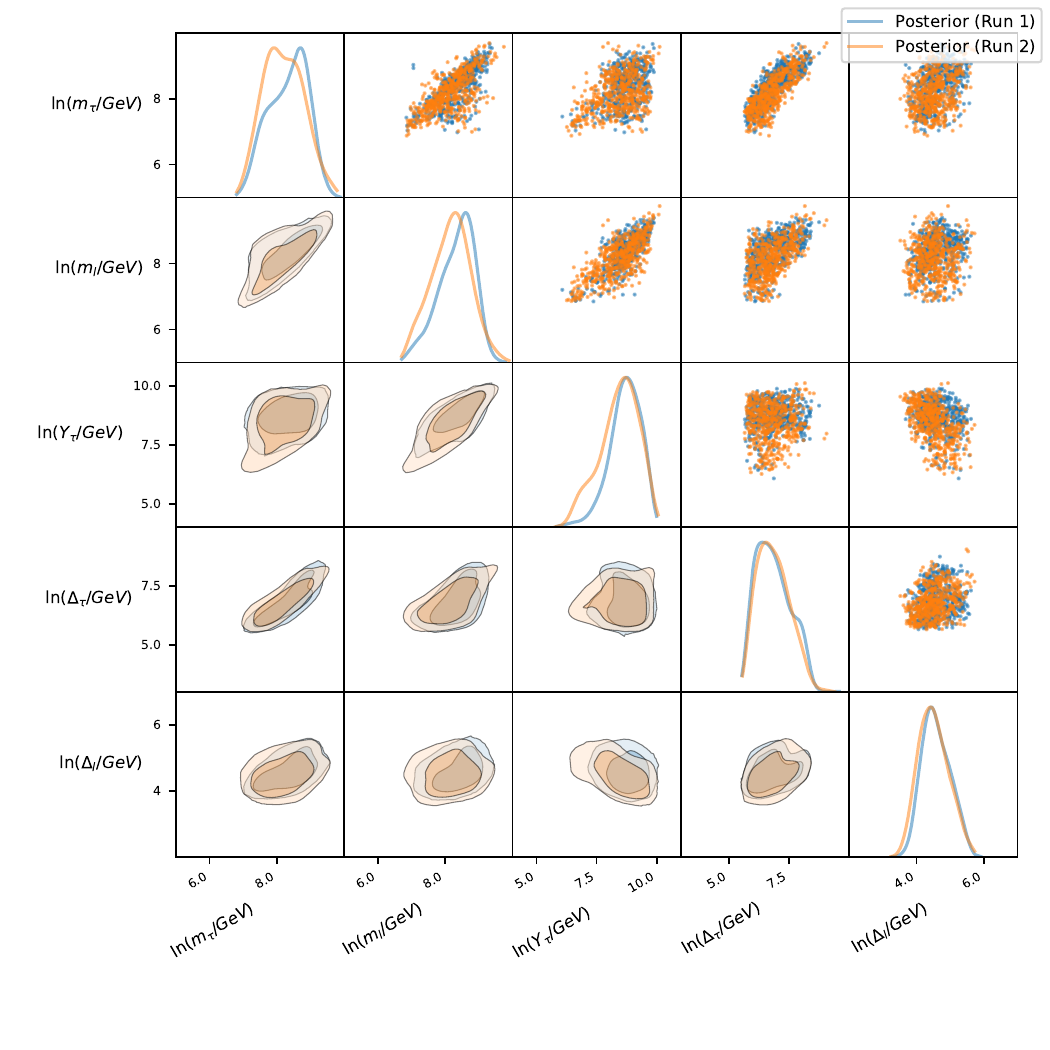}
\caption{1D and 2D marginalised posteriors for the tau lepton sector parameters in the LM4DCHM$^{5-5-5}_{14-10}$ found in two different runs with $4000$ live points.}
\label{fig:14-10_tau_posteriors}
\end{figure}

\Cref{fig:14-10_gauge_posteriors,fig:14-10_top_posteriors,fig:14-10_bottom_posteriors,fig:14-10_tau_posteriors} show the posterior distributions found in the two 4000-point scans of the LM4DCHM$^{5-5-5}_{14-10}$. There is seen to be good agreement in the posteriors between the two scans, especially in the lepton sector. The evidences found in each scan,
\begin{align}
    \ln(\mathcal{Z})_{\text{Run 1}} &= -35.91 \pm 0.07, \nonumber\\
    \ln(\mathcal{Z})_{\text{Run 2}} &= -36.67 \pm 0.07,
\end{align}
are in slight tension with each other, but are acceptably consistent given the same considerations as for the LM4DCHM$^{5-5-5}_{5-5}$ above.

\clearpage